\documentclass[11pt]{article}% добавляется twoside для двусторонней печати

\usepackage[T1]{fontenc}
\usepackage[centertags]{amsmath}
\usepackage{amsfonts}
\usepackage{amssymb}
\usepackage[pdftex]{hyperref}
\usepackage{graphicx}
\usepackage{graphbox}
\usepackage[numbers,sort&compress]{natbib}% упорядочивает ссылки

\usepackage{paperinitial}% Установка параметров страницы

\paperinitialization{15mm}{15mm}{15mm}{15mm}{2pt}{10pt}

\DeclareMathOperator{\re}{Re}

\DeclareMathOperator{\tg}{tg}
\DeclareMathOperator{\arctg}{arctg}
\def\res{\mathop{\mathrm{res}}}

\newcommand{\lan}{\langle}
\newcommand{\ran}{\rangle}

\newcommand{\e}{\varepsilon}
\newcommand{\vf}{\varphi}
\newcommand{\vk}{\varkappa}
\newcommand{\s}{\sigma}

\newcommand{\al}{\alpha}
\newcommand{\be}{\beta}
\newcommand{\ga}{\gamma}
\newcommand{\Ga}{\Gamma}
\newcommand{\de}{\delta}
\newcommand{\De}{\Delta}

\newcommand{\la}{\lambda}

\newcommand{\ups}{\upsilon}

\newcommand{\spx}{\mathbf{x}}
\newcommand{\spy}{\mathbf{y}}

\newcommand{\spk}{\mathbf{k}}
\newcommand{\spe}{\mathbf{e}}

\newcommand{\spb}{\mathbf{b}}

\begin{document}%\selectlanguage{english}
\allowdisplaybreaks[4]% позволяет переносить многострочные формулы
\frenchspacing% уменьшение пробелов после запятых
\setlength{\unitlength}{1pt}% устанавливает единицу длины в окружении picture

\title{{\Large\textbf{Probability of radiation of twisted photons by cold relativistic particle bunches}}}

\date{}

\author{O.V. Bogdanov${}^{1),2)}$\thanks{E-mail: \texttt{bov@tpu.ru}},\; P.O. Kazinski${}^{1)}$\thanks{E-mail: \texttt{kpo@phys.tsu.ru}},\; and G.Yu. Lazarenko${}^{1)}$\thanks{E-mail: \texttt{laz@phys.tsu.ru}}\\[0.5em]
{\normalsize ${}^{1)}$ Physics Faculty, Tomsk State University, Tomsk 634050, Russia}\\
{\normalsize ${}^{2)}$ Tomsk Polytechnic University, Tomsk 634050, Russia}}

\maketitle

\begin{abstract}

The probability to record a twisted photon produced by a cold relativistic particle bunch of charged particles is derived. The radiation of twisted photons by such particle bunches in stationary electromagnetic fields and in propagating electromagnetic waves is investigated. Several general properties of both incoherent and coherent contributions to the radiation probability of twisted photons are established. It is shown that the incoherent radiation by bunches of particles traversing normally an isotropic dispersive medium (the edge, transition, and Vavilov-Cherenkov radiations) and by bunches moving in a helical undulator does not depend on the azimuthal distribution of particles in the bunch and is the same as for round bunches. As for planar undulators, the incoherent radiation by particle bunches is the same as for the bunches symmetric under reflection with respect to the axis of a twisted photon detector. At high energies of recorded twisted photons, this property is universal and holds for the forward incoherent radiation by any cold relativistic particle bunch. The coherent radiation of twisted photons by such particle bunches obeys the property that we call the addition rule. This rule provides a simple means to describe the properties of coherent radiation of twisted photons. Furthermore, the strong addition rule is established for the coherent radiation by sufficiently long helical bunches. The use of this rule allows one to elaborate superradiant pure sources of twisted photons. The coherent radiation by helical bunches is considered for the edge, transition, and Vavilov-Cherenkov processes and for particles moving in undulators and plane laser waves with circular polarization. In these cases, the sum rules are deduced for the total probability to record a twisted photon and for the projection of the total angular momentum per photon. The explicit expressions for both incoherent and coherent interference factors are derived for several simple bunch profiles.

\end{abstract}

\section{Introduction}

Nowadays there are several experimental techniques to generate twisted photons by bunches of relativistic charged particles moving in undulators or striking metal foils \cite{EAllaria08,HKDXMHR,BHKMSS,KatohSRexp,Hemsing14hel,HemsingTR12,HemStuXiZh14,Rubic17}. Surprisingly, a rigorous quantum theory for the probability of radiation of twisted photons by bunches of particles has barely been developed (see, however, \cite{Kaminer16,BKb}). The main goal of this paper is to fill this gap. As long as particle bunches contain a huge number of particles, it is, of course, impossible to find exactly the probability of radiation of twisted photons by such bunches. Nevertheless, in many practical applications, a good zeroth order approximation is a bunch of charged particles moving along parallel trajectories, i.e., a cold particle bunch. Therefore, we restrict our considerations to the study of properties of radiation of twisted photons by such bunches evolving in external electromagnetic fields. The configuration of the electromagnetic fields is assumed to be of such a form that the parallel transport of trajectories is a symmetry of the particle equations of motion on the scale of the bunch waist. In this particular case, we derive the general formula for the probability to record a twisted photon produced by the particle bunch and investigate some its general properties. Several interesting applications are also considered.

In the soft X-ray spectral range and below, the twisted photons can be generated indirectly by the usual means using spatial light modulators, spiral phase plates, diffraction gratings with dislocations, etc (see for review \cite{KnyzSerb,PadgOAM25,Roadmap16,AndBabAML,TorTorTw,AndrewsSLIA}). However, in order to produce hard twisted photons or elaborate a bright source of them, one has to employ direct processes involving the radiation by bunches of relativistic particles \cite{HKDXMHR,BHKMSS,KatohSRexp,Hemsing14hel,HemsingTR12,HemStuXiZh14,Rubic17,BKL2,ABKT,EpJaZo,BKL3,BKL4,HemMar12,HemMarRos11,RibGauNin14,HemRos09,IvSerZay,SasMcNu,TaHaKa,KatohPRL,TairKato,JenSerprl,JenSerepj,Ivanov11,AfanMikh,BordKN,Kaminer16,BKb}. For example, it was shown in \cite{RibGauNin14,HemMar12,HemMarRos11,HemRos09,HemStuXiZh14,HKDXMHR,HemsingTR12} that specially designed beams of charged particles (the helical beams) can be used as a superradiant source of twisted photons when they are launched to an undulator or hit a metal foil. The generated twisted photons can be used further in various applications (see, e.g., \cite{KnyzSerb,PadgOAM25,Roadmap16,AndBabAML,TorTorTw,AndrewsSLIA,MHSSF13,PesFriSur15,AfSeSol18,SGACSSK}) or recorded by a specially designed detector of twisted photons. At the present moment, there are elaborated several methods to record twisted photons in the X-ray range \cite{PSTP19} and below \cite{LPBFAC,BLCBP,SSDFGCY,LavCourPad,RGMMSCFR}. Thus the knowledge of properties of radiation of twisted photons by bunches of particles can also be employed for development of new methods of diagnostics of the bunch structure (see \cite{HemsingTR12,HemRos09,HLarocque16} for the examples of such techniques).

Having derived in Sec. \ref{Gen_Form} the general formulas for probability to record a twisted photon by a cold relativistic particle bunch, we deduce several general properties of twisted photon radiation produced by such bunches. In particular, in Sec. \ref{Incoh_Inter_Fact}, we establish a universal property that, at high energies of recorded twisted photons, the main contribution to probability comes from the even Fourier harmonics of the density of particles in the bunch with respect to the azimuth angle. The incoherent contribution to the radiation probability by charged particles in a planar wiggler possesses the same property for arbitrary energies. So, in these cases, the radiation of twisted photons looks as if it is created by a particle bunch symmetric with respect to reflection in the detector axis.

Another interesting general property of radiation of twisted photons by particle bunches is the addition rule. We prove it in Secs. \ref{Stat_Field}, \ref{Generali}. The addition rule says that the $k$-th harmonic of the Fourier series of the particle density with respect to the azimuth angle gives the contribution to the coherent radiation of twisted photons only with the projection of the total angular momentum $m=j+k$, where $j$ is the projection of the total angular momentum of a twisted photon radiated by one particle moving along the bunch center. Notice that the particle density with dominant harmonics $\pm k$ looks as a helix with $k$ branches. For sufficiently long helical bunches, the coherent production of twisted photons is possible. In that case, we establish in Sec. \ref{Stat_Field} the strong addition rule asserting that, for the coherent radiation of helical bunches, the spectrum of twisted photons over $m$ produced by one particle moving along the bunch center is shifted by $n$, where $n$ is the signed number of the coherent harmonic. These rules refine and generalize the observations made in \cite{RibGauNin14,HemMar12,HemMarRos11,HemRos09,HemStuXiZh14,HKDXMHR,HemsingTR12} on the properties of radiation produced by helical bunches. The use of these rules allows one to elaborate the bright pure sources of twisted photons. At present, there are several developed techniques to manipulate the particle beam profiles in order to generate a coherent radiation (see, e.g., \cite{RibGauNin14,HemMar12,HemMarRos11,HemRos09,HemStuXiZh14,HKDXMHR,HemsingTR12,EAllaria08,Hemsing14hel,Rubic17,XiStup09,Hemsing7516,BGarcia16,Stupakov09,TLiu19,AMak19,EASeddon17,Hemsing19,ACurcio19,PRRibic19}).

In Sec. \ref{Sum_Rules}, we generalize the sum rules for the incoherent radiation obtained in \cite{BKb} to the case of an arbitrary motion of the particle bunch. Furthermore, in Sec. \ref{Stat_Field}, considering the radiation produced by sufficiently long helical bunches, we simplify the general formula for the probability to record a twisted photon and find the sum rules for the total radiation probability and for the projection of the total angular momentum per photon. These formulas interpolate between the cases of completely coherent and completely incoherent radiation. In Secs. \ref{Expl_Expr_Incoh}, \ref{Expl_Expr_Coh}, we derive the explicit expressions for the incoherent and coherent interference factors for simple bunch profiles and investigate their asymptotics.

We use the conventions and notation adopted in \cite{BKL2}. In particular, we use the system of units such that $\hbar=c=1$ and $e^2=4\pi\al$, where $\al$ is the fine structure constant.

\section{General formulas}\label{Gen_Form}

\subsection{Radiation probability}

Let us derive the general formula for the probability to record a twisted photon in the radiation produced by a bunch of charged particles moving along parallel trajectories. The amplitude of radiation of a twisted photon is characterized by the helicity $s$, the projection $m$ of the total angular momentum onto the detector axis, the projection $k_3$ of the photon momentum onto this axis, and the absolute value of the perpendicular photon momentum component $k_\perp$. Under the parallel translations by the $4$-vector $\de^\mu$ in the spacetime, this amplitude is transformed as \cite{BKL3,BKb}
\begin{equation}\label{parall_transp}
    A(\de;s,m,k_3,k_\perp)=\sum_{n=-\infty}^\infty e^{ik_0\de^0-ik_3\de^3}j^*_{m-n}(k_\perp\de_+,k_\perp\de_-)A(0;s,n,k_3,k_\perp),
\end{equation}
where $A(0;s,m,k_3,k_\perp)$ is the initial amplitude. Henceforth, the spatial indices are risen and lowered with the help of the Euclidean metric $\de_{ij}$. We also work in the coordinate system adapted to the detector of twisted photons with the axis directed along the unit vector $\spe_3$ (see for details \cite{BKL2}). The unit vectors $\{\spe_1,\spe_2,\spe_3\}$ of this coordinate system constitute a right-hand triple. Besides, we use the notation \cite{BKL2}
\begin{equation}
    \de_\pm:=\de_1\pm i\de_2,
\end{equation}
and analogously for other vectors. Notice that formula \eqref{parall_transp} is a general one and can be employed when the twisted photon is created by a quantum current. In that case, $A(0;s,m,k_3,k_\perp)$ depends on the form of the radiating particle wave packet and takes the quantum recoil into account. If the bunch of $N$ particles moving along parallel trajectories radiates, then the amplitude is written as
\begin{equation}\label{amplitude_bunch}
    A(s,m,k_3,k_\perp)=\sum_{a=1}^N\sum_{n=-\infty}^\infty e^{ik_0\de^0_a-ik_3\de^3_a}j^*_{m-n}(k_\perp\de^a_+,k_\perp\de^a_-)A(0;s,n,k_3,k_\perp),
\end{equation}
where $\de^\mu_a\equiv\de^\mu(\spb_a)$ and $\spb_a$ specifies the particle position in the bunch with respect to the center of the bunch (for example, with respect to some distinguished particle) at a given instant of laboratory time. The explicit expressions for $\de^\mu(\spb_a)$ will be given below.

With the typical number of particles in the bunch, $N\sim 10^{10}$, it is impossible to prepare the bunch of particles with given $\spb_a$. Therefore, we suppose that the initial particle positions in the bunch can be specified only with a certain probability density $\rho(\spb_1,\cdots,\spb_N)$:
\begin{equation}
    \int d\spb_1\cdots d\spb_N\rho(\spb_1,\cdots,\spb_N)=1.
\end{equation}
We also assume that the particles are identical and their initial positions in the bunch are uncorrelated, viz.,
\begin{equation}\label{prob_distr}
    \rho(\spb_1,\cdots,\spb_N)=\prod_{a=1}^N\rho(\spb_a),\qquad\int d\spb\rho(\spb)=1.
\end{equation}
The observables are the expectation values with respect to the distribution $\rho(\spb_1,\cdots,\spb_N)$. In particular, the average density of particles at the point $\spb$ equals
\begin{equation}
    \int d\spb_1\cdots d\spb_N\rho(\spb_1,\cdots,\spb_N) \sum_{a=1}^N\de(\spb-\spb_a)=N\rho(\spb),
\end{equation}
where $\rho(\spb)$ is the one-particle distribution.

The probability to record a twisted photon by the detector is the amplitude squared (for more details, see \cite{BKL2}). Averaging the square of amplitude \eqref{amplitude_bunch} over the distribution \eqref{prob_distr}, we arrive at the natural splitting of the average radiation probability into the incoherent and coherent contributions
\begin{equation}\label{prob_rad_incoh_coh}
\begin{split}
    dP_\rho(s,m,k_3,k_\perp)=&\,N\sum_{n,n'}f_{m-n,m-n'}A(0;s,n,k_3,k_\perp)A^*(0;s,n',k_3,k_\perp)+\\
    &+N(N-1)\sum_{n,n'}\vf_{m-n}\vf^*_{m-n'} A(0;s,n,k_3,k_\perp)A^*(0;s,n',k_3,k_\perp),
\end{split}
\end{equation}
where the incoherent, $f_{mn}$, and coherent, $\vf_m$, interference factors have been introduced
\begin{equation}\label{interfer_factor}
\begin{split}
    f_{mn}&:=\int d\spb\rho(\spb)j^*_m\big(k_\perp\de_+(\spb),k_\perp\de_-(\spb)\big) j_n\big(k_\perp\de_+(\spb),k_\perp\de_-(\spb)\big),\\
    \vf_m&:=\int d\spb\rho(\spb) e^{ik_0\de^0(\spb)-ik_3\de^3(\spb)} j^*_m\big(k_\perp\de_+(\spb),k_\perp\de_-(\spb)\big).
\end{split}
\end{equation}
It is clear that
\begin{equation}\label{symm_prop}
    f_{mn}=f^*_{nm}=(-1)^{m+n}f_{-n,-m}.
\end{equation}
The last equality follows from relation [(A5), \cite{BKL2}]. So long as $\rho(\spb)$ is nonnegative, $f_{mn}$ is Hermitian positive-definite.

In fact, it is assumed in \eqref{prob_rad_incoh_coh} that the initial state of the bunch of particles is described by the mixed state characterized by the probability distribution \eqref{prob_distr} and almost the same initial momenta. If $A(0;s,n,k_3,k_\perp)$ describes the amplitude of twisted photon production by a quantum current, then the use of formula \eqref{prob_rad_incoh_coh} corresponds to the approximation implying that the quantum correlations of particles in the bunch are negligible and the particle wave packets are obtained from each other by a parallel transport. In itself, the coherent contribution in \eqref{prob_rad_incoh_coh} can be employed for a semiclassical description of radiation generated by the wave packet of one particle provided the momenta of modes constituting the wave packet are almost the same and, in radiating twisted photons, the quantum recoil is negligibly small.

\subsection{Particular cases}

Let us find the explicit expressions for $\de^\mu(\spb)$ in two particular cases. Consider, at first, the motion of a bunch of identical particles in the field that is invariant with respect to translations with the parameter
\begin{equation}\label{transl_statio}
    \de^\mu=(\de^0,\boldsymbol{\de}),\qquad \boldsymbol{\xi}\boldsymbol{\de}=0,
\end{equation}
in the region of spacetime occupied by the bunch. Here $\boldsymbol{\xi}=(0,\sin\theta,\cos\theta)$ characterizes the direction along which the field changes rapidly on the bunch scale. The typical example of the field, which is invariant with respect to translations \eqref{transl_statio}, is the stationary electromagnetic field of undulator with the axis directed along $\boldsymbol{\xi}$, provided  the transverse size of the particle bunch is sufficiently small. Another example of such a field is a homogeneous medium with nontrivial permittivity. In this case, the vector $\boldsymbol{\xi}$ is a normal to the vacuum-medium interface at the points where the bunch enters and exits the medium.

When the bunch moves in the region where the external field is absent, i.e., before the interaction with this field, the trajectory $y^\mu(x^0)$ of the particle in the bunch with the coordinate $\spb$ with respect to the center of the bunch at a given instant of laboratory time can be obtained by translation \eqref{transl_statio} from the central trajectory $x^\mu(x^0)$ as
\begin{equation}\label{traj_parall}
    \spx(x^0)\rightarrow \spy(x^0)=\spx(x^0)+\boldsymbol{\de}=\spx(0)+\boldsymbol{\be}x^0+\boldsymbol{\de},\qquad x^0\rightarrow y^0(x^0)=x^0+\de^0,
\end{equation}
where $\boldsymbol{\be}$ is the particle velocity before interaction with the field. Setting $\spx(0)=0$, $\spy(x^0)=\spb$, and $y^0(x^0)=0$, and using \eqref{transl_statio}, we obtain the system of equations for $\de^\mu$. Its solution is given by
\begin{equation}\label{transl_statio1}
    \de^0=-\frac{\boldsymbol{\xi} \spb}{\boldsymbol{\xi}\boldsymbol{\be}}=-\frac{b_2\tg\theta+b_3}{\be_2\tg\theta+\be_3},\qquad \de_+=b_+-\be_+\frac{\boldsymbol{\xi}\spb}{\boldsymbol{\xi}\boldsymbol{\be}},\qquad \de_3=\frac{\be_2 b_3-\be_3 b_2}{\be_2\tg\theta+\be_3}\tg\theta.
\end{equation}
In virtue of the assumption about the structure of the external field, transformations \eqref{transl_statio1} are the symmetry transformations and map the solution of particle equations of motion to the solution of equations of motion in the given field. Recall  we suppose that the interaction between particles in the bunch is negligible on the radiation formation scale.

Another interesting class of translations is comprised by translations with the parameter $\de^\mu$ subject to the condition
\begin{equation}\label{transl_wave}
    \xi_\mu \de^\mu=0,\qquad \xi^\mu=(1,\boldsymbol{\xi}),\quad \boldsymbol{\xi}^2=1,
\end{equation}
where one can choose $\boldsymbol{\xi}=(0,\sin\theta,\cos\theta)$. We suppose that the external field varies weakly under the translations $\de^\mu$ in the region of spacetime occupied by the bunch. For example, such a field corresponds to the field of a plane (in the vicinity of the bunch) electromagnetic wave propagating along the vector $\boldsymbol{\xi}$. The joint solution of \eqref{traj_parall}, \eqref{transl_wave} is
\begin{equation}\label{transl_wave_1}
    \de^0=\frac{\boldsymbol{\xi} \spb}{1-\boldsymbol{\xi}\boldsymbol{\be}},\qquad\boldsymbol{\de}=\spb +\boldsymbol{\be}\frac{\boldsymbol{\xi}\spb}{1-\boldsymbol{\xi}\boldsymbol{\be}}.
\end{equation}
These translations are the symmetry transformations of the particle equations of motion in the region of localization of the bunch.

Let us consider the coherent interference factor \eqref{interfer_factor} in more detail. Introducing
\begin{equation}
    \spk_\perp:=k_\perp(\sin\psi,\cos\psi,0),\qquad \spk:=(\spk_\perp,k_3),
\end{equation}
and employing integral representation [(A8), \cite{BKL2}] of the Bessel functions, we find
\begin{equation}\label{interfer_factor_coh}
    \vf_m=\int_{-\pi}^\pi\frac{d\psi}{2\pi} e^{im\psi}\int d\spb\rho(\spb)e^{ik_0\de^0(\spb)-i\spk\boldsymbol{\de}(\spb)}=\int_{-\pi}^\pi\frac{d\psi}{2\pi}e^{im\psi}\int d\spb\rho(\spb)e^{ik_0[\de^0(\spb)-\mathbf{n}\boldsymbol{\de}(\spb)]},
\end{equation}
where $\mathbf{n}:=\spk/k_0$. In the cases discussed above, $\de^\mu$ is a linear function of $\spb$. Hence,
\begin{equation}
    k_0[\de^0(\spb)-\mathbf{n}\boldsymbol{\de}(\spb)]=k_0(\de^0_i-\mathbf{n}\boldsymbol{\de}_i)b^i.
\end{equation}
If $\rho(\spb)$ is an infinitely differentiable function absolutely integrable with all its derivatives, then by the Riemann-Lebesgue lemma the Fourier transform appearing in \eqref{interfer_factor_coh} tends to zero faster than any inverse power of
\begin{equation}\label{expont_expr}
    k^2_0\sum_i(\de^0_i-\mathbf{n}\boldsymbol{\de}_i)^2
\end{equation}
as \eqref{expont_expr} tends to infinity.

As for translations \eqref{transl_statio1}, condition \eqref{expont_expr} is equivalent to $k_0\rightarrow\infty$ for any $\mathbf{n}$ when $\boldsymbol{\xi}\boldsymbol{\be}\neq0$. In this case, $\vf_m$ tends to zero faster than any power of $k_0^{-1}$ as $k_0\rightarrow\infty$. As regards the translations \eqref{transl_wave_1}, condition \eqref{expont_expr} is equivalent to $k_0\rightarrow\infty$ only for the vectors $\mathbf{n}\neq \boldsymbol{\xi}$. In this case, $\vf_m$ tends to zero faster than any power of $k_0^{-1}$ as $k_0\rightarrow\infty$ provided $\boldsymbol{\xi}$ does not coincide with $\mathbf{n}$ for any azimuth angle $\psi$, i.e., the vector $\boldsymbol{\xi}$ does not belong to the cone directed along the detector axis $\spe_3$ with the aperture $2\arctg n_\perp$. We shall show below that the incoherent interference factor \eqref{interfer_factor} decreases as a power at large $k_0$. Therefore, the incoherent contribution dominates in \eqref{prob_rad_incoh_coh} for sufficiently large energies of detected photons.

\paragraph{Forward radiation.}

Further, we restrict our considerations to the case when one can set $\theta=\{0,\pi\}$ and $\be_\perp:=|\be_+|=0$. We call such a situation as forward radiation. It follows from the explicit expressions for the interference factors \eqref{interfer_factor} that, in the case \eqref{transl_statio1}, it is necessary to assume that $\be_\perp\De\theta\ll\be_3$ and
\begin{equation}
\begin{alignedat}{2}
    k_0\frac{\s_\perp\De\theta}{\be_3}&\ll1,&\qquad k_0\frac{\s_3\be_\perp\De\theta}{\be_3^2}&\ll1,\\
    k_0 n_3\s_\perp\De\theta&\ll1,&\qquad k_0\frac{n_3\s_3\be_\perp\De\theta}{\be_3}&\ll1,\\
    k_0 \frac{n_\perp\s_\perp\be_\perp\De\theta}{\be_3}&\ll1,&\qquad k_0\frac{n_\perp\s_3\be_\perp}{\be_3}&\ll1,
\end{alignedat}
\end{equation}
where $\s_3$ and $\s_\perp$ are the typical sizes of the bunch along the axis $\spe_3$ and in the transverse directions, respectively, $\be_\perp$ is the deviation of $\be_\perp$ from zero, and $\De\theta\ll1$ is the deviation of the angle $\theta$ from the values specified above. As for the incoherent interference factor, one needs only the fulfillment of the conditions on the third line and $\be_\perp\De\theta\ll\be_3$. Notice that the fulfillment of the first condition on the first line implies the fulfillment of the first condition on the third line.

In the ultrarelativistic regime, $\ga\gg1$, in the region where the main part of radiation is concentrated, the following estimates hold (see, e.g., \cite{BKL2})
\begin{equation}\label{estimates}
    |\beta_\pm|\equiv\be_\perp\lesssim \vk/\gamma,\qquad|\beta_3|\approx1,\qquad |n_\perp|\lesssim \vk/\gamma,\qquad n_3\approx1,
\end{equation}
where $\vk:=\max(1,K)$, $K:=\lan\be_\perp\ran\ga$, and $\lan\be_\perp\ran$ is a typical value of $\be_\perp$ during the evolution. Therefore, for the coherent interference factor, we have to demand $\De\theta\ll1$ and
\begin{equation}
    k_0\s_\perp\De\theta\ll1,\qquad k_0\be_\perp\s_3\De\theta\ll1,\qquad k_0\frac{\vk}{\ga}\be_\perp\s_3\ll1.
\end{equation}
For the incoherent interference factor, we obtain the conditions
\begin{equation}
    k_0\frac{\vk}{\ga}\be_\perp\s_\perp\De\theta\ll1,\qquad k_0\frac{\vk}{\ga}\be_\perp\s_3\ll1,
\end{equation}
and $\De\theta\ll1$.

As regards the field configurations invariant with respect to translations \eqref{transl_wave_1}, one can set $\theta=\{0,\pi\}$ and $\be_\perp=0$, provided
\begin{equation}\label{forw_conds}
    \frac{\be_\perp\De\theta}{1-\zeta\be_3}\ll1,
\end{equation}
and
\begin{equation}\label{forw_conds1}
\begin{alignedat}{2}
    k_0\frac{\s_\perp\De\theta}{1-\zeta\be_3}&\ll1,&\qquad k_0\frac{\s_3\be_\perp\De\theta}{(1-\zeta\be_3)^2}&\ll1,\\
    k_0 \frac{n_3\be_3\s_\perp\De\theta}{1-\zeta\be_3}&\ll1,&\qquad k_0\frac{n_3\be_3\s_3\be_\perp\De\theta}{(1-\zeta\be_3)^2}&\ll1,\\
    k_0 \frac{n_\perp\s_\perp\be_\perp\De\theta}{1-\zeta\be_3}&\ll1,&\qquad k_0\frac{n_\perp\s_3\be_\perp}{1-\zeta\be_3}&\ll1,
\end{alignedat}
\end{equation}
where $\zeta=\cos\theta=\pm1$. For the incoherent interference factor, it is only necessary the fulfillment of \eqref{forw_conds} and the conditions on the third line of \eqref{forw_conds1}. The first condition on the third line of \eqref{forw_conds1} follows from the first condition in \eqref{forw_conds1}.

In the ultrarelativistic regime, the cases $\zeta=\pm1$ have to be considered separately. For $\zeta=1$, in case of the coherent interference factor, we obtain
\begin{equation}\label{forw_conds_ew}
    \frac{2\be_\perp\ga^2}{1+\be_\perp^2\ga^2}\De\theta\ll1,
\end{equation}
and
\begin{equation}
    k_0\frac{2\ga^2}{1+\be_\perp^2\ga^2}\s_\perp\De\theta\ll1,\qquad k_0\frac{4\be_\perp\ga^4}{(1+\be_\perp^2\ga^2)^2}\s_3\De\theta\ll1,\qquad \frac{k_0}{\vk} \frac{2\be_\perp\ga}{1+\be_\perp^2\ga^2}\s_3\ll1,
\end{equation}
where we have taken into account that the main part of radiation is concentrated at $n_\perp\sim1/(\gamma\vk)$ \cite{BKL4}. As for the incoherent interference factor, one needs the fulfillment of \eqref{forw_conds_ew} and
\begin{equation}
    \frac{k_0}{\vk} \frac{2\be_\perp\ga}{1+\be_\perp^2\ga^2}\s_\perp\De\theta\ll1,\qquad \frac{k_0}{\vk} \frac{2\be_\perp\ga}{1+\be_\perp^2\ga^2}\s_3\ll1.
\end{equation}
For $\zeta=-1$, in case of the coherent interference factor, we have
\begin{equation}\label{forw_conds_ew-1}
    \be_\perp \De\theta/2\ll1,
\end{equation}
and
\begin{equation}
    k_0 \s_\perp\De\theta/2\ll1,\qquad k_0 \be_\perp\s_3\De\theta/4\ll1,\qquad k_0 \vk\be_\perp \s_3/(2\ga)\ll1.
\end{equation}
In this case, the main part of radiation is concentrated at $n_\perp\sim\vk/\gamma$ \cite{BKL4}. As for the incoherent interference factor, the conditions when one can take $\theta=\pi$ and $\be_\perp=0$ become
\begin{equation}
    k_0 \vk\be_\perp \s_\perp\De\theta/(2\ga)\ll1,\qquad k_0 \vk\be_\perp \s_3/(2\ga)\ll1,
\end{equation}
provided \eqref{forw_conds_ew-1} is satisfied.

\section{Incoherent interference factor}\label{Incoh_Inter_Fact}

\subsection{Sum rules}\label{Sum_Rules}

Let us consider separately the properties of the incoherent interference factor. As was discussed above, at large energies of detected photons, the main contribution to probability \eqref{prob_rad_incoh_coh} comes from the first term depending on the incoherent interference factor,
\begin{equation}\label{incoh_prob}
    dP^{nc}_\rho(s,m,k_3,k_\perp):= N\sum_{n,n'}f_{m-n,m-n'}A(0;s,n,k_3,k_\perp)A^*(0;s,n',k_3,k_\perp).
\end{equation}
In this case, the following sum rule holds \cite{BKb}:
\begin{equation}
    dP^{nc}_\rho(s,k_3,k_\perp):=\sum_{m=-\infty}^\infty dP^{nc}_\rho(s,m,k_3,k_\perp)=N\sum_{m=-\infty}^\infty dP_1(s,m,k_3,k_\perp),
\end{equation}
where $dP_1(s,m,k_3,k_\perp)$ is the probability of twisted photon radiation by one charged particle. This sum rule is a consequence of the relation
\begin{equation}\label{sum_rule1}
    \sum_{m=-\infty}^\infty f_{m-n,m-n'}=\de_{n,n'}.
\end{equation}
Furthermore, the another sum rule is satisfied:
\begin{multline}\label{dJ3_sumrule}
    dJ_{3\rho}^{nc}(s,k_3,k_\perp):=\sum_{m=-\infty}^\infty mdP_\rho(s,m,k_3,k_\perp)=NdJ_{31}(s,k_3,k_\perp)+\\ +N\re\sum_{n=-\infty}^\infty A(0;s,n,k_3,k_\perp)A^*(0;s,n-1,k_3,k_\perp)\int d\spb \rho(\spb)k_\perp\de_+(\spb),
\end{multline}
where $dJ_{31}^{nc}(s,k_3,k_\perp)$ is the average value of projection of the total angular momentum of twisted photons produced by one charged particle.

If the last term in \eqref{dJ3_sumrule} vanishes, then
\begin{equation}\label{ang_mom_sr}
    dJ_{3\rho}^{nc}(s,k_3,k_\perp)=NdJ_{31}(s,k_3,k_\perp),\qquad \ell_\rho^{nc}(s,k_3,k_\perp):=\frac{dJ_{3\rho}^{nc}(s,k_3,k_\perp)}{ dP^{nc}_\rho(s,k_3,k_\perp)}=\ell_1(s,k_3,k_\perp).
\end{equation}
The last term in \eqref{dJ3_sumrule} can be zero due to a special symmetry of the bunch or due to properties of the one-particle radiation amplitude of twisted photons (see \cite{BKb}). For example, the last term in \eqref{dJ3_sumrule} vanishes with good accuracy for the forward undulator radiation in the dipole regime, for the forward radiation of ideal helical and planar wigglers, and for the radiation of particles moving strictly along the axis of a twisted photon detector.

\subsection{Forward radiation}

Now we turn to the case $\theta=\{0,\pi\}$ and $\be_\perp=0$. In both cases \eqref{transl_statio1} and \eqref{transl_wave_1}, we have
\begin{equation}\label{interfer_factor_incoh}
    f_{mn}=\int d\spb\rho(\spb)j^*_m(k_\perp b_+,k_\perp b_-) j_n(k_\perp b_+,k_\perp b_-).
\end{equation}
Let $\spb_\perp:=(b_1,b_2)$ and
\begin{equation}\label{rho_bar}
    \bar{\rho}(\spb_\perp):=\int db_3\rho(\spb).
\end{equation}
We suppose that this integral converges absolutely and uniformly with respect to $\spb_\perp$. Then, if $\rho(\spb)$ is an infinitely differentiable function, then $\bar{\rho}(\spb_\perp)$ is also infinitely differentiable.

The function $\bar{\rho}(\spb_\perp)$ can be developed as a Fourier series with respect to the azimuth angle
\begin{equation}\label{Fourier_ser}
    \bar{\rho}(\spb_\perp)=\sum_{n=-\infty}^\infty c_n(b_\perp/\s_\perp) e^{in\psi},\qquad c^*_n=c_{-n},
\end{equation}
where $\s_\perp$ specifies the transverse size of the bunch. As long as $\bar{\rho}(\spb_\perp)\geqslant0$, the matrix $c_{n-k}$ is Hermitian positive-definite. It follows from the explicit expression for $c_n$ in terms of $\bar{\rho}(\spb_\perp)$ that $\bar{\rho}(\spb_\perp)$ is infinitely differentiable at $\spb_\perp=0$, only if
\begin{equation}\label{Fourier_coeff}
    c_n(r)=r^{|n|}\bar{c}_n(r^2),
\end{equation}
where $\bar{c}_n(y)$ are infinitely smooth functions for $y\in[0,\infty)$. The normalization condition for the probability density \eqref{prob_distr} becomes
\begin{equation}\label{norm_cond}
    2\pi\s_\perp^2\int_0^\infty dr rc_0(r)=1.
\end{equation}
Substituting \eqref{Fourier_ser} into \eqref{interfer_factor_incoh}, we obtain
\begin{equation}\label{inter_fact_incoh}
    f_{mn}(x)=2\pi\s_\perp^2\int_0^\infty drrc_{m-n}(r)J_m(xr)J_n(xr),
\end{equation}
where $x:=k_\perp\s_\perp$. In the particular case of axially symmetric bunches, when $c_n=0$ for $n\neq0$, we arrive at the incoherent interference factor,
\begin{equation}\label{incoh_int_fact_round}
    f_{mn}(x)=\de_{mn}f_m(x),
\end{equation}
studied in \cite{BKb}. Notice that the round bunches are created, for example, in the electron-positron collider VEPP-2000, Novosibirsk \cite{RPPPDG2018}. In a general case, $f_{mm}(x)=f_m(x)$.

\subsection{Examples}

Let us consider several interesting particular cases. The first one is the forward radiation of an ideal helical wiggler with the helicity $\chi=\pm1$ or the forward radiation produced by charged particles in the laser wave with circular polarization and smooth envelope. In this case, the radiation of twisted photons by one particle obeys the selection rule $m=\chi n$ \cite{Rubic17,KatohPRL,TaHaKa,SasMcNu,TairKato,BKL4,BKL2,BKL3,BHKMSS,BKb,Hemsing14hel,KatohSRexp}, where $n$ is the harmonic number of radiation. Hence, the probability of incoherent radiation produced by a bunch of particles at the $n$-th harmonic takes the form
\begin{equation}\label{incoh_hel_und}
    dP_\rho^{nc}(s,m,k_3,k_\perp)=Nf_{m-\chi n,m-\chi n}dP_1(s,\chi n,k_3,k_\perp).
\end{equation}
The incoherent interference factor appearing in this expression,
\begin{equation}
    f_{m-\chi n,m-\chi n}(x)=f_{m-\chi n}(x),
\end{equation}
is determined only by the harmonic $c_0$ of the Fourier series \eqref{Fourier_ser}. To put it differently, in this case the probability of incoherent radiation of twisted photons does not depend on the angular distribution of particles in the bunch and is the same as for an axially symmetric bunch \cite{BKb}. The total probability \eqref{prob_rad_incoh_coh} simplifies to
\begin{equation}\label{prob_rad_tot_hw}
    dP_\rho(s,m,k_3,k_\perp)=\big[Nf_{m-\chi n}+N(N-1)|\vf_{m-\chi n}|^2\big]dP_1(s,\chi n,k_3,k_\perp).
\end{equation}
The coherent contribution depends only on the harmonic $\tilde{c}_{m-\chi n}$ of the Fourier series of the probability density $\rho(\spb)$ with respect to the azimuth angle. This is a particular manifestation of the addition rule that will be discussed in Sec. \ref{Coh_Int_Fact}.

The same properties are inherent to the radiation produced by charged particles moving strictly along the detector axis. The radiation by one particle moving along the detector axis consists of the twisted photons with $m=0$ \cite{BKL3}. Therefore, in this case the incoherent radiation and the total radiation have the form \eqref{incoh_hel_und}, \eqref{prob_rad_tot_hw}, respectively, with $n=0$. Such a situation is realized, for example, for the edge radiation \cite{BlNord37,Nords37,AkhBerQED,BolDavRok,WeinbergB.12,AkhShul,Bord.1,GinzbThPhAstr,GinzTsyt} of a charged particle that stops instantaneously or starts to move along the detector axis, for a charged particle falling normally onto an ideally conducting plate, and for a charged particle moving along the detector axis in an isotropic medium  -- the Vavilov-Cherenkov (VCh) and transition radiations \cite{GinzbThPhAstr,GinzTsyt}. The detailed investigation of radiation of twisted photons by charged particles moving in an isotropic dispersive medium will be given elsewhere \cite{BKLm}.

%uniform bunch
\begin{figure}[tp]
\centering
\includegraphics*[align=c,width=0.49\linewidth]{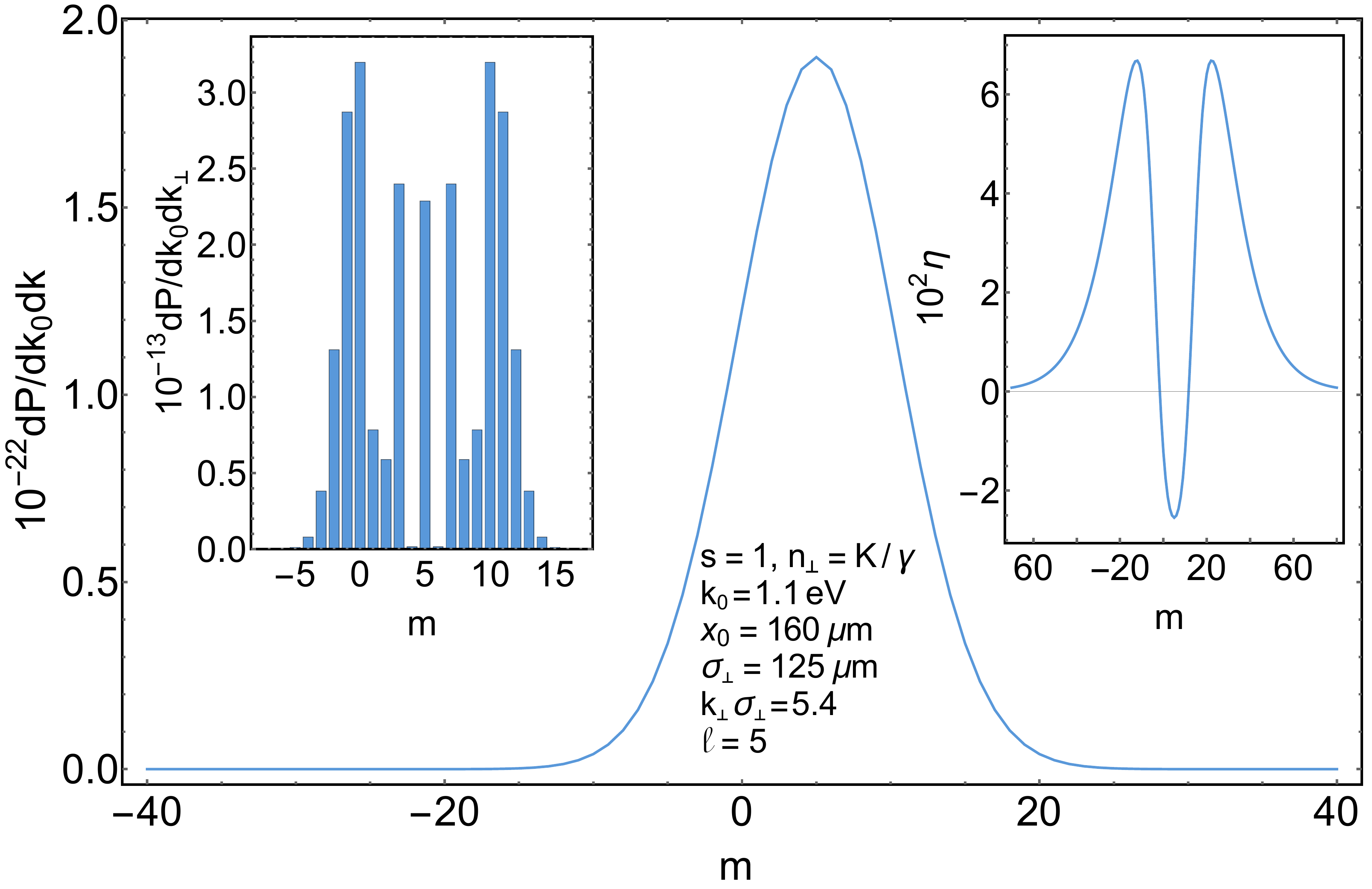}\;
\includegraphics*[align=c,width=0.49\linewidth]{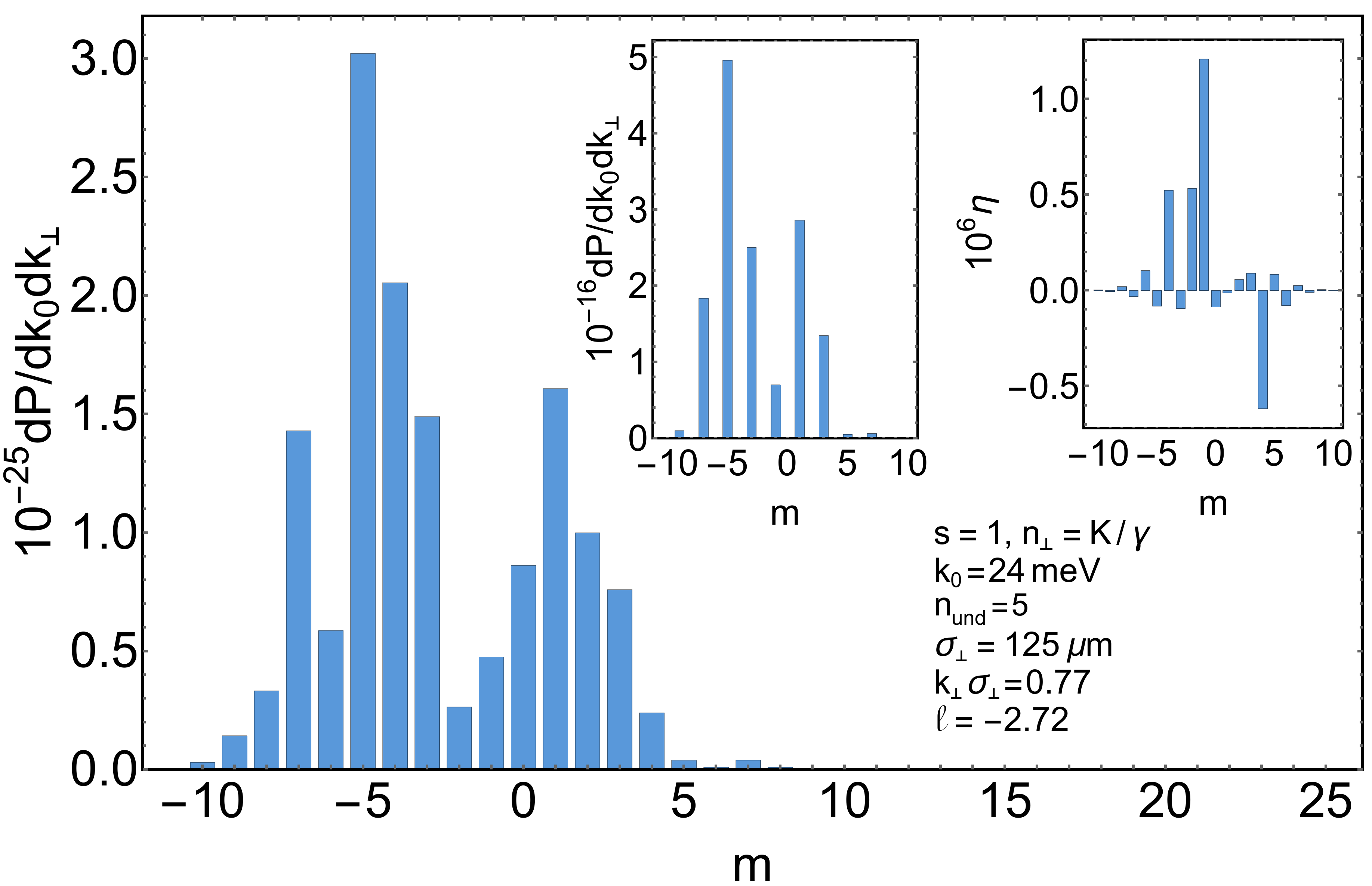}
\caption{{\footnotesize The probability of the forward radiation of twisted photons for a uniform Gaussian bunch of electrons in the wiggler at the fifth harmonic. The one-particle probability distribution has the form \eqref{Gaussian_distr}, \eqref{Gauss_Fourier_coeff} with $\s_3=150$ $\mu$m (duration $0.5$ ps), $\s_\perp=125$ $\mu$m, $h_x=0.1$, $h_y=0.05$, the eccentricity is $0.9$, which corresponds to $\e_x=1.77$ and $\e_y=0.77$. The number of particles in the bunch is $3\times10^8$. The coherent contribution to radiation is negligible in this case. The left inset is the one-particle radiation probability distribution over $m$. The right inset shows $\eta:=(dP_\rho-dP_{\rho_e})/dP_\rho$, where $dP_\rho$ is the radiation probability for the initial particle bunch and $dP_{\rho_e}$ is the radiation probability for the same bunch but where only even Fourier harmonics with respect to the azimuth angle are retained. On the left panel: The helical wiggler is considered. The Lorentz factor is $\ga=700$, the undulator strength parameter is $K=5$, and the undulator period is $\la_0=10$ cm. The particle moving long the center of the bunch enters the undulator at the offset $x_0=160$ $\mu$m from the undulator axis. We see on the right inset that, at large photon energies, the contribution of odd Fourier harmonics \eqref{Gauss_Fourier_coeff} is suppressed. As follows from the sum rule \eqref{ang_mom_sr}, the projection of the total angular momentum per photon does not depend on the bunch profile and is the same as for radiation created by one particle moving along an ideal helix with the axis coinciding with the detector axis. On the right panel: The planar wiggler is considered. The Lorentz factor is $\ga=100$, the undulator strength parameter is $K=5$, and the undulator period is $\la_0=10$ cm. The contribution of odd Fourier harmonics of \eqref{Gaussian_distr} is negligible. The projection of the total angular momentum per photon does not depend on the bunch profile and is the same as for the radiation created by one particle.}}
\label{uniform bunch_fig}
\end{figure}

In the case of the forward radiation of a planar wiggler, it was shown in \cite{BKL2} that the radiation of twisted photons obeys the selection rule: $m+n$ is an even number, where $n$ is the harmonic number. Substituting this condition into \eqref{incoh_prob}, it is easy to see that the incoherent radiation of twisted photons by a bunch of particles in this wiggler is determined solely by the even Fourier harmonics $c_k$. The incoherent radiation is such as if it is created by a particle bunch symmetric with respect to reflection $\spb_\perp\rightarrow-\spb_\perp$.

In the case of the forward undulator radiation in the dipole regime, the main part of radiation consists of the twisted photons with $m=\pm1$ \cite{BKL2}. Then
\begin{multline}
    dP_\rho^{nc}(s,m,k_3,k_\perp)=N\Big\{f_{m-1,m-1}dP_1(s,1,k_3,k_\perp)+f_{m+1,m+1}dP_1(s,-1,k_3,k_\perp)+\\ +2\re\big[f_{m-1,m+1}A(0;s,-1,k_3,k_\perp)A^*(0;s,1,k_3,k_\perp)\big]\Big\}.
\end{multline}
Therefore, the incoherent forward radiation in the dipole regime is determined only by the Fourier harmonics $c_0$ and $c_{\pm2}$.

\subsection{Asymptotics}

Consider the asymptotics of \eqref{inter_fact_incoh} at large and small $x$. It is clear that
\begin{equation}
    f_{00}(0)=1;\qquad f_{mn}(0)=0,\; m\neq0,\,n\neq0.
\end{equation}
Expanding the product of two Bessel functions in a Taylor series \cite{GrRy}, we deduce
\begin{equation}\label{smlx_expan}
    f_{mn}(x)=2\pi\s_\perp^2\sum_{k=0}^\infty\frac{(-1)^k}{k!}\Big(\frac{x}{2}\Big)^{m+n+2k}\frac{(m+n+k+1)_k}{\Ga(m+k+1)\Ga(n+k+1)} \int_0^\infty drr^{m+n+2k+1}c_{m-n}(r),
\end{equation}
where $(n)_k$ is the Pochhammer symbol and it is assumed that $c_k(r)$ tend to zero at infinity faster than any power of $r^{-1}$.

In order to find the asymptotics of $f_{mn}(x)$ at large $x$, we shall employ the procedure expounded, for example, in \cite{ParKam01,Wong,KalKaz3}. It is convenient to consider the interference factor $f_{m,m-k}$ instead of $f_{mn}$. Then the Mellin transform of the product of two Bessel functions entering into \eqref{inter_fact_incoh} has the form \cite{PruBryMar2}
\begin{equation}
    \int_0^\infty drr^{\nu-1}J_m(xr)J_{m-k}(xr)=\Big(\frac{2}{x}\Big)^{\nu} \frac{\Ga(1-\nu)\Ga\big(m+(\nu-k)/2\big)}{2\Ga\big(1-(\nu+k)/2\big)\Ga\big(1-(\nu-k)/2\big)\Ga\big(1+m-(\nu+k)/2\big)},
\end{equation}
where $\re\nu<1$ and $\re\nu>k-2m$. We assume for a while that $k-2m<1/2$. In that case,
\begin{equation}\label{incoh_fact_M}
    f_{m,m-k}(x)=2\pi\s_\perp^2\int_C\frac{d\mu}{2\pi i} \Big(\frac{2}{x}\Big)^{\mu} \frac{\Ga(1-\mu)\Ga\big(m+(\mu-k)/2\big) M_k(2-\mu)}{2\Ga\big(1-(\mu+k)/2\big)\Ga\big(1-(\mu-k)/2\big)\Ga\big(1+m-(\mu+k)/2\big)},
\end{equation}
where the contour $C$ runs from below upwards in the strip $k-2m<\re\mu<1$ and
\begin{equation}\label{Mell_trans}
    M_k(\nu):=\int_0^\infty drr^{\nu-1}c_k(r),
\end{equation}
which is understood in the sense of analytic continuation for those $\nu$ where integral \eqref{Mell_trans} diverges (for details, see \cite{ParKam01,Wong,KalKaz3,GSh}). The asymptotic expansion of $f_{m,m-k}$ at large $x$ is obtained by shifting the integration contour in \eqref{incoh_fact_M} to the right with the account for singularities of the integrand. The singularities of the gamma function and its expansion in the vicinity of these singularities are well known. The singularities of the function $M_k(\nu)$ and its expansion near them can also be found explicitly \cite{ParKam01,Wong,KalKaz3,GSh}.

Let $\mu=\mu'+i\mu''$ on the contour $C$. The integral \eqref{incoh_fact_M} converges for $x>0$, provided
\begin{equation}
    \mu'<1/2+\e_k(\mu'),
\end{equation}
where $\e(\mu')$ is the decrease power of $M_k(2-\mu)$ on the contour $C$, viz.,
\begin{equation}
    |M_k(2-\mu)|_{C}<c|\mu''|^{-\e_k(\mu')},
\end{equation}
where $c$ is some constant. If $c_k(r)$ tend to zero at infinity faster than any power of $r^{-1}$ and are expandable in the Taylor series in the vicinity of $r=0$, then $\e_k(\mu')>0$ and the function $M_k(\nu)$ possesses the singularities in the form of simple poles at the points $\nu=-n$, $n=\overline{0,\infty}$, with the residues \cite{ParKam01,Wong,KalKaz3,GSh}
\begin{equation}
    \res_{\nu=-n}M_k(\nu)=c_k^{(n)}(0)/n!.
\end{equation}
One can shift the contour $C$ to the right so long as the integral over $\mu$ is converging. Further, we additionally assume that  $c_k(r)$ are infinitely differentiable and tend to zero with all their derivatives faster than any power of $r^{-1}$ at infinity. In that case, employing the Riemann-Lebesgue lemma, one can readily show that $|M_k(2-\mu)|$ vanishes faster than any power of $|\mu''|^{-1}$ as $|\mu''|\rightarrow\infty$, i.e., $\e_k(\mu')=\infty$. Then the integration contour $C$ can be moved to the right up to infinity.

\begin{figure}[tp]
\centering
\includegraphics*[align=c,width=0.49\linewidth]{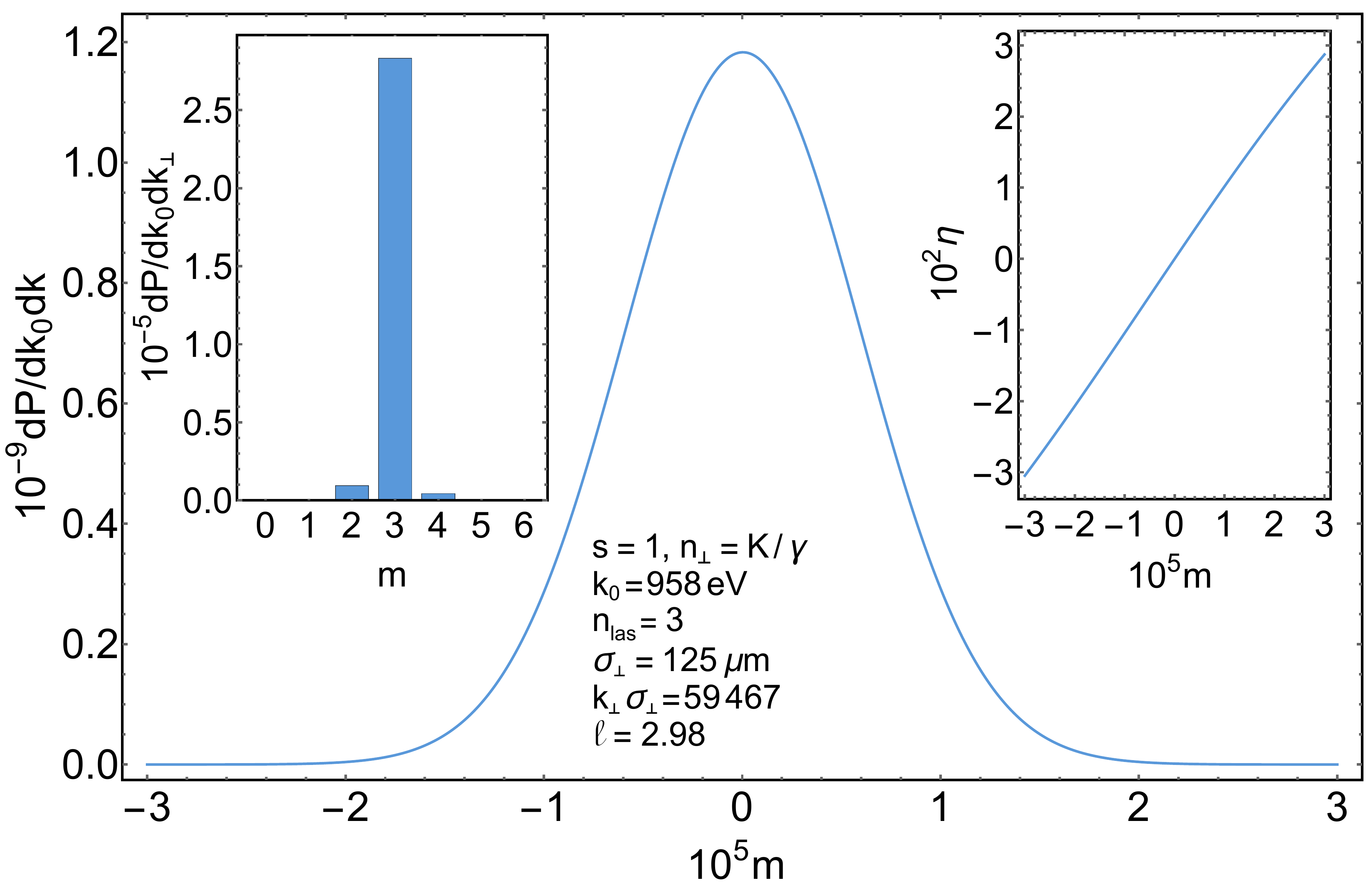}\;
\caption{{\footnotesize The same as in Fig. \ref{uniform bunch_fig} but for radiation created by the electron bunch with $\ga=100$ in  head-on collision with the circularly polarized electromagnetic wave produced by Ti:Sa laser with the intensity $10^{22}$ W/cm${}^2$, the photon energy $\Omega=1.53$ eV, and the amplitude envelope $f_0\sin^4(\Omega\xi/(2N))$ with $N=20$ (see the notation in \cite{BKL4}). The third harmonic is considered. The contribution of coherent radiation is negligible. The contribution of odd Fourier harmonics \eqref{Gauss_Fourier_coeff} is strongly suppressed. The projection of the total angular momentum per photon does not depend on the bunch profile and is the same as for the radiation created by one particle.}}
\label{uniform bunch1_fig}
\end{figure}

In virtue of uniqueness of the analytic continuation, the asymptotic expansion obtained above is also valid in the domain $k-2m\geqslant1/2$. We will not derive the complete asymptotic expansion and give only the leading asymptote for large $x$. In the case $c_k(0)\neq0$, we deduce
\begin{equation}\label{incoh_asym1}
\begin{split}
    f_{m,m-k}&\simeq2\pi\s_\perp^2\Big\{\frac{\cos\tfrac{\pi k}{2}}{\pi x}\int_0^\infty drc_k(r)+\frac{k(2m-k)}{2\pi x^2}\sin\frac{\pi k}{2}\Big[\int_0^1 \frac{dr}{r}(c_k(r)-c_k(0))+\int_1^\infty\frac{dr}{r}c_k(r)+\\
    &+\frac{c_k(0)}{2}\big[\ln\frac{x^2}{4}+2\psi(2)-\psi(1+m-k/2)-\psi(m-k/2)-\psi(k/2)-\psi(-k/2)\big] \Big] \Big\}+\cdots,
\end{split}
\end{equation}
where $\psi(x):=\Ga'(x)/\Ga(x)$. If $c_k(0)=0$, then
\begin{equation}\label{incoh_asym2}
    f_{m,m-k}\simeq2\pi\s_\perp^2\Big\{\frac{\cos\tfrac{\pi k}{2}}{\pi x}\int_0^\infty drc_k(r) +\frac{k(2m-k)}{2\pi x^2}\sin\frac{\pi k}{2}\int_0^\infty\frac{dr}{r}c_k(r)\Big\} +\cdots.
\end{equation}
As we see, the contribution of even Fourier harmonics $c_k$ to the probability of incoherent radiation dies out slower than the contribution of odd harmonics for large $x$. Therefore, at large energies of recorded twisted photons, the radiation will look as if it is created by a particle bunch invariant under reflection $\spb_\perp\rightarrow-\spb_\perp$ (see Figs. \ref{uniform bunch_fig}, \ref{uniform bunch1_fig}, \ref{beam_helix_fig}).

\paragraph{Generating function.}

It is not difficult to obtain the generating function for the incoherent interference factors
\begin{equation}\label{gener_func}
    \sum_{m=-\infty}^\infty t^mf_{m,m-k}(x)=(-1)^k(-t)^{k/2}\bar{\vf}_k\Big(x\frac{1-t}{\sqrt{-t}}\Big),
\end{equation}
where the principal branches of multivalued functions are taken and
\begin{equation}\label{coh_inter_red}
    \bar{\vf}_k(y):=2\pi\s_\perp^2\int_0^\infty drrc_k(r)J_k(yr).
\end{equation}
The function $\bar{\vf}_k(y)$ is related to the coherent interference factor \eqref{coh_inter_fact2} for helical bunches. In particular, for $t=1$, we obtain the sum rule \eqref{sum_rule1}. Formula \eqref{gener_func} implies the integral representation
\begin{equation}\label{gener_func_inv}
    f_{m,m-k}(x)=(-1)^k\int_{C'}\frac{dt}{2\pi i}t^{-m-1}(-t)^{k/2}\bar{\vf}_k\Big(x\frac{1-t}{\sqrt{-t}}\Big),
\end{equation}
where the integration is carried out along the closed contour $C'$ going in the positive direction around the point $t=0$ and lying in the ring of analyticity of the integrand.

\subsection{Explicit expressions}\label{Expl_Expr_Incoh}

Let us obtain the explicit expressions for $f_{mn}(x)$ for some typical bunch profiles. These profiles allow one to simulate the real particle bunches and to demonstrate explicitly the main properties of the interference factors.

a) Uniform distribution
\begin{equation}\label{uniform_distr}
    c_k(r)=\frac{\al_k}{\pi\s_\perp^2} r^{|k|}\theta(1-r),
\end{equation}
where $\al_k=\al^*_{-k}\in \mathbb{C}$ are some constant such that the matrix $c_{k-n}(r)$ is Hermitian positive-definite. The normalization condition leads to $\al_0=1$. The form of distribution \eqref{uniform_distr} is the simplest one that complies with \eqref{Fourier_coeff} and vanishes for $r>1$. When $m\geqslant0$, it follows from \eqref{inter_fact_incoh} that
\begin{equation}
    f_{m,m-k}=\left\{
                     \begin{array}{ll}
                       \dfrac{\al_k}{1+k}\big[J_{m}(x)J_{m-k}(x)-J_{m+1}(x)J_{m-k-1}(x)\big], & k\geqslant0; \\[0.8em]
                       \dfrac{\al_k}{1-k}\big[J_{m}(x)J_{m-k}(x)-J_{m-1}(x)J_{m-k+1}(x)\big], & k<0,
                     \end{array}
                   \right.
\end{equation}
where $x=k_\perp\s_\perp$. As regards the negative values of $m$, the interference factor can be obtained with the aid of the symmetry property \eqref{symm_prop}. For small $x$ and $m\geqslant0$, we have
\begin{equation}
    f_{m,m-k}=\left\{
                     \begin{array}{ll}
                       \dfrac{\al_k(-1)^{m+k}}{(k+1)m!(k-m)!}\Big(\dfrac{x}{2}\Big)^{k}+\cdots, & k\geqslant m; \\[0.8em]
                       \dfrac{\al_k}{(m+1)!(m-k)!}\Big(\dfrac{x}{2}\Big)^{2m-k}+\cdots, & m\geqslant k\geqslant0; \\[0.8em]
                       \dfrac{\al_k}{m!(m-k+1)!}\Big(\dfrac{x}{2}\Big)^{2m-k}+\cdots, & k<0.
                     \end{array}
                   \right.
\end{equation}
For $x\gg\max(1,|m|,|k|)$, $m\geqslant0$, the asymptotics reads
\begin{equation}\label{incoh_uni}
    f_{m,m-k}\simeq\al_k\left\{
                     \begin{array}{ll}
                       \dfrac{2\cos\frac{\pi k}{2}}{\pi x (1+k)}+\dfrac{(2m-k)\sin\frac{\pi k}{2}-(-1)^m\cos(\frac{\pi k}{2}+2x)}{\pi x^2}+\cdots, & k\geqslant 0; \\[0.8em]
                       \dfrac{2\cos\frac{\pi k}{2}}{\pi x (1-k)}-\dfrac{(2m-k)\sin\frac{\pi k}{2}+(-1)^m\cos(\frac{\pi k}{2}+2x)}{\pi x^2}+\cdots, & k<0.
                     \end{array}
                   \right.
\end{equation}
Notice that, in the case at issue, the assumptions that were used to derive general formulas \eqref{incoh_asym1}, \eqref{incoh_asym2} for the asymptotic expansion of the interference factor at large $x$ are not satisfied. Therefore, \eqref{incoh_uni} does not coincide with \eqref{incoh_asym1}, \eqref{incoh_asym2}. General formulas \eqref{incoh_asym1}, \eqref{incoh_asym2} reproduce only the powerlike part of the asymptotics. Nevertheless, just as for \eqref{incoh_asym1}, \eqref{incoh_asym2}, the contribution of the odd Fourier harmonics is suppressed in comparison with the contribution of the even Fourier harmonics at large $x$.

b) Gaussian bunch
\begin{equation}\label{Gauss_distr}
    c_k(r)=\frac{\al_k}{2\pi\s_\perp^2} r^{|k|}e^{-r^2/2}.
\end{equation}
The normalization condition leads to $\al_0=1$. As in the previous example, distribution \eqref{Gauss_distr} is the simplest expression complying with \eqref{Fourier_coeff} and possessing the Gaussian profile. For $m\geqslant0$, we have
\begin{equation}
    f_{m,m-k}=\al_k\left\{
                     \begin{array}{ll}
                       \dfrac{2^{k-m}x^{2m-k}}{(m-k)!}F(m-k/2+1/2,m-k/2+1;m-k+1,2m-k+1;-2x^2), & k\geqslant0; \\[0.8em]
                       \dfrac{2^{-m}x^{2m-k}}{m!}F(m-k/2+1/2,m-k/2+1;m+1,2m-k+1;-2x^2), & k<0.
                     \end{array}
                   \right.
\end{equation}
For small $x$, $m\geqslant0$, we come to
\begin{equation}
    f_{m,m-k}=\al_k\left\{
                     \begin{array}{ll}
                       \dfrac{(-1)^{m+k}k!}{m!(k-m)!}x^{k}+\cdots, & k\geqslant m; \\[0.8em]
                       \dfrac{2^{k-m}}{(m-k)!}x^{2m-k}+\cdots, & m\geqslant k\geqslant0; \\[0.8em]
                       \dfrac{2^{-m}}{m!}x^{2m-k}+\cdots, & k<0.
                     \end{array}
                   \right.
\end{equation}
Recall that, for negative $m$, the expression for $f_{mn}$ can be obtained by means of relation \eqref{symm_prop}. The asymptotics for $x\gg\max(1,|m|,|k|)$, $m\geqslant0$, reads
\begin{equation}
    f_{m,m-k}\simeq\al_k\left\{
                     \begin{array}{ll}
                       \dfrac{2^{(k-1)/2}}{\Ga\big((1-k)/2\big) x}+\dfrac{2^{k/2-1}(k-2m)}{\Ga(-k/2) x^2}+\cdots, & k\geqslant 0; \\[0.8em]
                       \dfrac{2^{-(k+1)/2}}{\Ga\big((1+k)/2\big) x}+\dfrac{2^{-k/2-1}(k-2m)}{\Ga(-k/2) x^2}+\cdots, & k<0.
                     \end{array}
                   \right.
\end{equation}
This expansion coincides with general formula \eqref{incoh_asym2}.

\begin{figure}[tp]
\centering
\includegraphics*[align=c,width=0.7\linewidth]{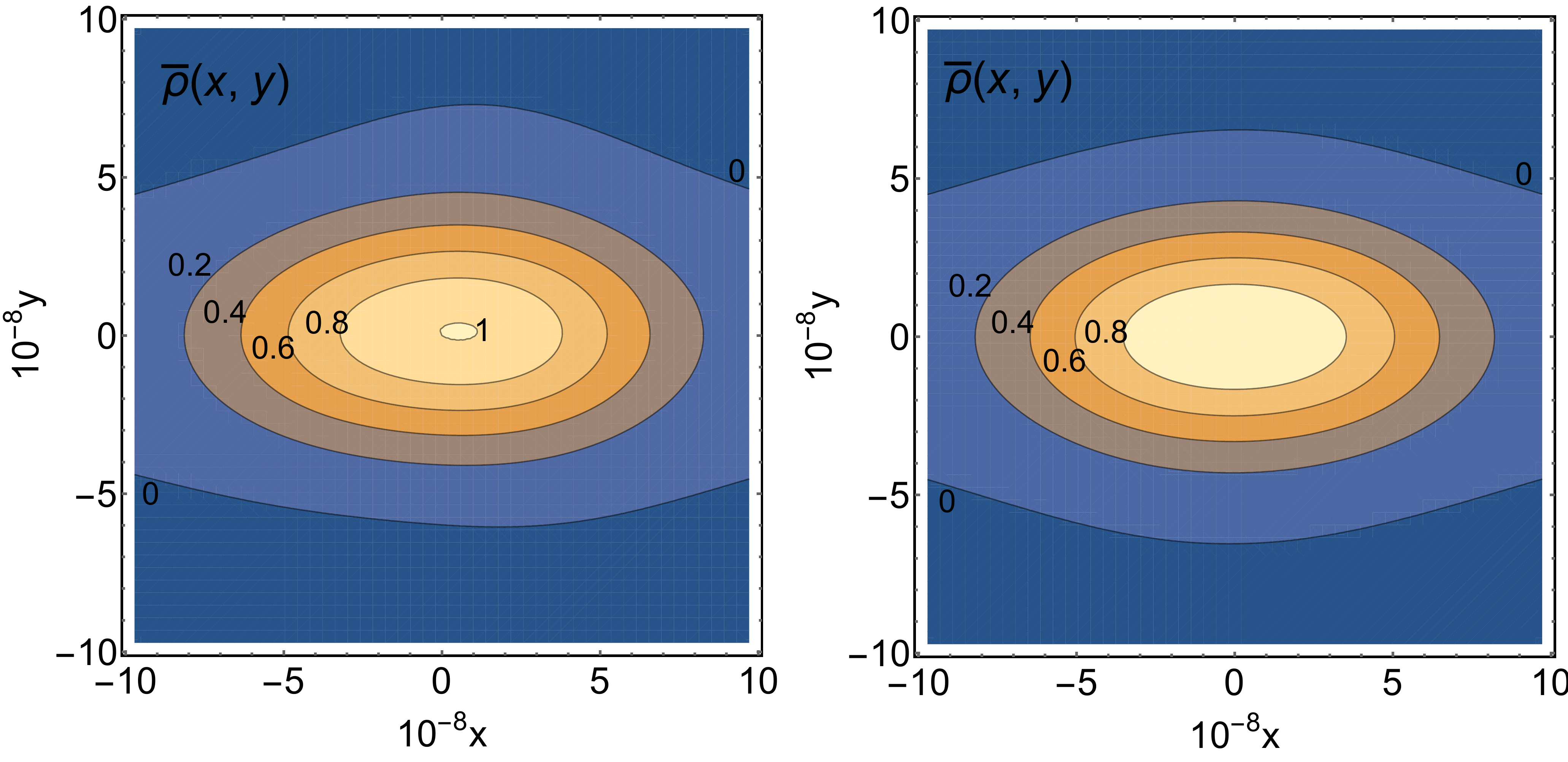}
\caption{{\footnotesize On the left panel: The integrated one-particle probability density $2\pi\s_\perp^2\bar{\rho}(\mathbf{b}_\perp)$ \eqref{rho_bar} for \eqref{Gauss_distr} with the parameters taken as in Fig. \ref{uniform bunch_fig}. The lengths are measured in the Compton wavelengths. On the right panel: The same as on the left panel but $\bar{\rho}(\spb_\perp)$ contains only even Fourier harmonics. It is clear that the symmetry property $\bar{\rho}(-\spb_\perp)=\bar{\rho}(\spb_\perp)$ is fulfilled.}}
\label{beam_helix_fig}
\end{figure}

To be more specific, we consider the Gaussian bunch of the form
\begin{equation}\label{Gaussian_distr}
    \rho(\spb)=\frac{1}{(2\pi)^{3/2}\s_3\s_x\s_y}e^{-b_3^2/(2\s_3^2)-(b_x-\s_x h_x)^2/(2\s_x^2)-(b_y-\s_y h_y)^2/(2\s_y^2)}.
\end{equation}
Hence,
\begin{equation}
    \bar{\rho}(\spb_\perp)=\frac{1}{2\pi\s_x\s_y}e^{-(b_x-\s_x h_x)^2/(2\s_x^2)-(b_y-\s_y h_y)^2/(2\s_y^2)}.
\end{equation}
The shift $h_{x,y}$ of the bunch center leads to the appearance of nonzero odd harmonics in the Fourier series \eqref{Fourier_ser}. Introducing the notation
\begin{equation}
\begin{gathered}
    \s^{-2}_\perp:=\frac{1}{2}(\s_x^{-2}+\s_y^{-2}),\qquad \e_{x}^2:=\frac{\s^2_{x}}{\s^2_\perp}=\frac12\Big(1+\frac{\s_x^2}{\s_y^2}\Big),\qquad \e_{y}^2:=\frac{\s_{y}^2}{\s^2_\perp}=\frac12\Big(1+\frac{\s_y^2}{\s_x^2}\Big),\\
    \e_{x}^{-2}+\e_{y}^{-2}=2,
\end{gathered}
\end{equation}
and employing the relation
\begin{equation}
    e^{z\cos\psi}=\sum_{k=-\infty}^\infty I_k(z)e^{ik\psi},
\end{equation}
where $I_n(z)$ are the modified Bessel functions of the first kind, the Fourier coefficients \eqref{Fourier_ser} can easily be found
\begin{equation}
    c_n=\frac{e^{-b_\perp^2/(2\s_\perp^2)}}{2\pi\e_x\e_y\s_\perp^2}e^{-h_\perp^2/2}\sum_{k=-\infty}^\infty I_k\Big(\frac{b^2_\perp}{4\s_\perp^2}(\e_y^{-2}-\e_x^{-2})\Big) I_{n-2k}\Big(\frac{b_\perp}{\s_\perp}(h_x^2/\e_x^2+h_y^2/\e_y^2)^{1/2}\Big)e^{i(n-2k)\phi},
\end{equation}
where $\phi=\arg(h_x/\e_x-ih_y/\e_y)$. Further, we assume that
\begin{equation}
    |\e_x-1|\ll1,\qquad|\e_y-1|\ll1,\qquad h_x^2\ll1,\qquad h_y^2\ll1.
\end{equation}
Then, in the leading order, we have
\begin{equation}\label{Gauss_Fourier_coeff}
\begin{split}
    c_0&=\frac{e^{-b_\perp^2/(2\s_\perp^2)}}{2\pi\s_\perp^2},\\
    c_n&=\frac{e^{-b_\perp^2/(2\s_\perp^2)}}{2\pi\s_\perp^2} \Big(\frac{b_\perp}{2\s_\perp}\Big)^{|n|}\bigg[\frac{e^{in\phi}}{|n|!}\Big(\frac{h_x^2}{\e_x^2}+\frac{h_y^2}{\e_y^2}\Big)^{|n|/2} +\frac{\de_{n,2k}}{|k|!}\Big(\frac{\e_y^{-2}-\e_x^{-2}}{2}\Big)^{|k|}\bigg],\quad n\neq0.
\end{split}
\end{equation}
These Fourier harmonics have the form \eqref{Gauss_distr}, and so the formulas above are applicable. The form of this bunch is presented in Figs. \ref{beam_helix_fig}, \ref{beam_helix_fig1}.

%form factors% incoherent
\begin{figure}[tp]
\centering
\includegraphics*[align=c,width=0.9\linewidth]{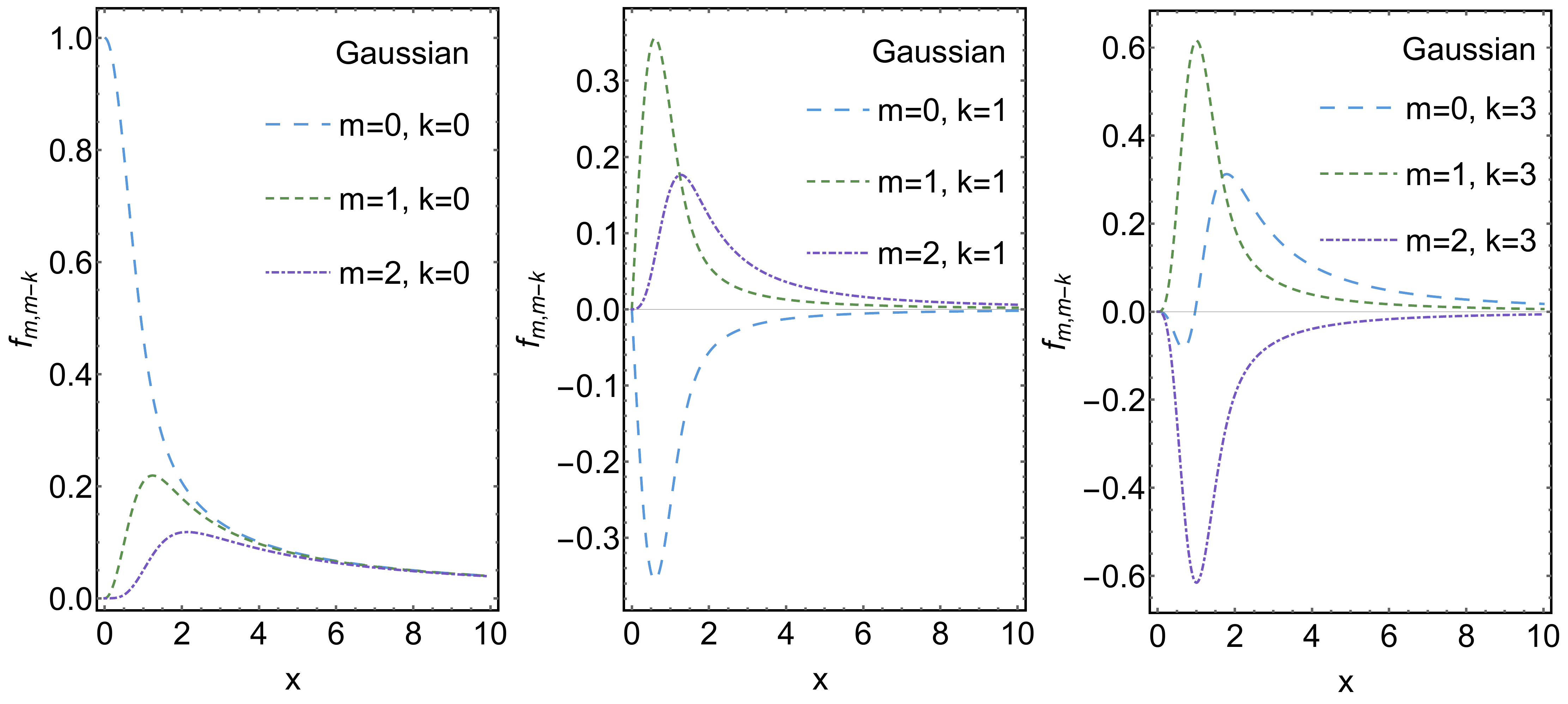}\;
\caption{{\footnotesize The normalized incoherent interference factor $f_{m,m-k}(x)/\alpha_k$ for the Gaussian bunch at different $m$ and $k$.}}
\label{form_factor_incoh_fig}
\end{figure}

c) Exponential profile
\begin{equation}\label{exp_distr}
    c_k(r)=\frac{\al_k}{2\pi\s_\perp^2}r^{|k|}e^{-r}.
\end{equation}
The normalization condition gives $\al_0=1$. The Fourier coefficients $c_k(r)$ are not of the form \eqref{Fourier_coeff}. The contribution of every $c_k(r)$ to series \eqref{Fourier_ser} is a function that is only $|k|$ times continuously differentiable at the point $\spb_\perp=0$. Therefore, the probability density $\bar{\rho}(\spb_\perp)$ is only a continuous function. Its first derivative has a discontinuity at the point $\spb_\perp=0$. We investigate such a profile because it is a limit of the generalized exponential profile (see below) and, unlike the latter, for this profile $f_{mn}$ can be found explicitly in terms of the known special functions.

For $m\geqslant0$, we have
\begin{equation}
    f_{m,m-k}=\al_k\left\{
                     \begin{array}{ll}
                       \dfrac{2^{1+k}\Ga(m+3/2) }{x^{k-2m}\sqrt{\pi}(m-k)!} F(m+\frac{3}{2},m-\frac{k-1}{2},m-\frac{k}{2}+1;m-k+1,2m-k+1;-4x^2), & k\geqslant0; \\[0.8em]
                       \dfrac{2^{1-k}\Ga(m-k+3/2)}{x^{k-2m}\sqrt{\pi}m!} F(m-k+\frac{3}{2},m-\frac{k-1}{2},m-\frac{k}{2}+1;m+1,2m-k+1;-4x^2), & k<0.
                     \end{array}
                   \right.
\end{equation}
For small $x$, $m\geqslant0$, we deduce
\begin{equation}\label{incoh_exp_sm}
    f_{m,m-k}=\al_k\left\{
                     \begin{array}{ll}
                       \dfrac{(-1)^{m+k}(2k+1)!}{m!(k-m)!}\Big(\dfrac{x}{2}\Big)^{k}+\cdots, & k\geqslant m; \\[0.8em]
                       \dfrac{2^{1+k}\Ga(m+3/2)}{\sqrt{\pi}(m-k)!}x^{2m-k}+\cdots, & m\geqslant k\geqslant0; \\[0.8em]
                       \dfrac{2^{1-k}\Ga(m-k+3/2)}{\sqrt{\pi}m!}x^{2m-k}+\cdots, & k<0.
                     \end{array}
                   \right.
\end{equation}
The asymptotics for $x\gg\max(1,|m|,|k|)$, $m\geqslant0$, takes the form
\begin{equation}\label{incoh_exp_lrg}
    f_{m,m-k}\simeq\al_k\left\{
                     \begin{array}{ll}
                       \dfrac{k!}{\pi x}\cos\frac{\pi k}{2}+\dfrac{(2m-k)k!}{2\pi x^2}\sin\frac{\pi k}{2}+\cdots, & k\geqslant 0; \\[0.8em]
                       \dfrac{|k|!}{\pi x}\cos\frac{\pi k}{2}-\dfrac{(2m-k)|k|!}{2\pi x^2}\sin\frac{\pi k}{2}+\cdots, & k<0.
                     \end{array}
                   \right.
\end{equation}
The leading terms of the expansion are in agreement with general formula \eqref{incoh_asym2}.

d) Generalized exponential profile
\begin{equation}\label{gen_exp_profile}
    c_k(r)=\frac{\al_k}{2\pi\s_\perp^2}r^{|k|}e^{-\sqrt{\la^2+r^2}},\quad\la>0.
\end{equation}
The normalization condition is reduced to
\begin{equation}\label{norm_cond_gexp}
    \al_0=\frac{e^\la}{1+\la}.
\end{equation}
The expression \eqref{gen_exp_profile} has the form \eqref{Fourier_coeff} and tends to zero at infinity as an exponent (for the wave packets with such an asymptote, see, e.g., \cite{KarlJHEPwp,BilBer17}). When $\la\rightarrow0$, the exponential profile considered above is reproduced.

Unfortunately, we did not succeed in finding a closed expression for the integral
\begin{equation}
    f_{m,m-k}=\al_k\int_0^\infty drr^{|k|+1}e^{-\sqrt{\la^2+r^2}}J_{m}(xr)J_{m-k}(xr)
\end{equation}
in terms of known special functions. The Mellin transform \eqref{Mell_trans} is written as \cite{GrRy}
\begin{equation}\label{Mell_trans_gexp}
    M_k(\nu)=\frac{\al_k}{2\sqrt{\pi}}\Ga\Big(\frac{|k|+\nu}{2}\Big)(2\la)^{(|k|+\nu+1)/2}K_{(|k|+\nu+1)/2}(\la).
\end{equation}
In particular, this formula and \eqref{norm_cond} entail the normalization condition \eqref{norm_cond_gexp}. Substituting \eqref{Mell_trans_gexp} into \eqref{incoh_fact_M}, we obtain the Mellin-Barnes representation for $f_{mn}$. Formulas \eqref{gener_func_inv}, \eqref{gen_exp_coh} provide another one integral representation for $f_{mn}$.

For small $x$, $m\geqslant0$, it follows from \eqref{smlx_expan}, \eqref{Mell_trans_gexp} that
\begin{equation}\label{incoh_gexp_sm}
    f_{m,m-k}=\frac{\al_k}{2\sqrt{\pi}}\left\{
                     \begin{array}{ll}
                       \dfrac{(-1)^{m+k}k!}{m!(k-m)!}\Big(\dfrac{x}{2}\Big)^{k}(2\lambda)^{k+3/2}K_{k+3/2}(\la)+\cdots, & k\geqslant m; \\[0.8em]
                       \dfrac{\Ga(m+1+(|k|-k)/2)}{(m-k)!}\Big(\dfrac{x}{2}\Big)^{2m-k}(2\lambda)^{m+3/2+(|k|-k)/2}K_{m+3/2+(|k|-k)/2}(\la)+\cdots, & k\leqslant m.
                     \end{array}
                   \right.
\end{equation}
The Macdonald functions entering into this expression are expressed through elementary functions. According to \eqref{incoh_asym2}, \eqref{Mell_trans_gexp}, the asymptotics for $x\gg\max(1,|m|,|k|)$, $m\geqslant0$, reads
\begin{equation}\label{incoh_gexp_lrg}
    f_{m,m-k}\simeq\frac{\al_k}{2\sqrt{\pi}}\Big[\frac{1}{x} \frac{(2\lambda)^{|k|/2+1}}{\Ga\big((1-|k|)/2\big)} K_{|k|/2+1}(\lambda)
    +\frac{|k|(2m-k)}{2 x^2} \frac{(2\lambda)^{(|k|+1)/2}}{\Ga(1-|k|/2)} K_{(|k|+1)/2}(\lambda)\Big]+\cdots.
\end{equation}
When $\la=0$, formulas \eqref{incoh_gexp_sm}, \eqref{incoh_gexp_lrg} turn into \eqref{incoh_exp_sm}, \eqref{incoh_exp_lrg}.

\section{Coherent interference factor}\label{Coh_Int_Fact}

As for the coherent interference factor, we investigate only the case $\theta=\{0,\pi\}$, $\be_\perp=0$ (the forward radiation) for helical bunches. We call the bunch helical if the one-particle probability distribution has the form
\begin{equation}\label{helical_bunch}
    \rho(\spb)=F(b_3)\rho_0(\spb)=F(b_3)\sum_{n=-\infty}^\infty c_n(b_\perp/\s_\perp)e^{in(\psi+2\pi\chi b_3/\de)},\qquad c^*_n=c_{-n},
\end{equation}
where $\psi$ is the azimuth angle, $\chi=\pm1$ specifies the handedness of the bunch, $\de$ is the helix pitch in the laboratory reference frame, and $\s_\perp$ defines the transverse size of the bunch. If $\rho_0(\spb)$ is an infinitely differentiable function, then $c_n(r)$ are of the form \eqref{Fourier_coeff}. The probability density $\rho_0(\spb)$ does not change under rotations around the detector axis by an angle of $\psi$ and simultaneous translations along this axis by $-\chi\psi\de/(2\pi)$. Taking
\begin{equation}
    \int db_3F(b_3)=1,
\end{equation}
the normalization condition for $\rho(\spb)$ is reduced to \eqref{norm_cond}. These bunches provide a generalization of axially symmetric bunches for which the radiation of twisted photons was considered in \cite{BKb}. Nowadays, there are several techniques to create such bunches \cite{RibGauNin14,HemMar12,HemMarRos11,HemRos09,HemStuXiZh14,HKDXMHR,HemsingTR12,CLiu16,XLZhu18,LBJu16}. The helical bunches were used in \cite{HKDXMHR} for generation of twisted photons by undulators at the first harmonic.

Notice that the incoherent radiation of twisted photons by helical bunches is described by formulas of the previous section with
\begin{equation}\label{rho_b}
    \bar{\rho}(\spb_\perp)=\sum_{n=-\infty}^\infty \tilde{F}(-2\pi\chi n/\de)c_n(b_\perp/\s_\perp)e^{in\psi},
\end{equation}
where
\begin{equation}
    \tilde{F}(p):=\int db_3e^{-ib_3p}F(b_3),\qquad \tilde{F}(0)=1,
\end{equation}
The total probability to record a twisted photon \eqref{prob_rad_incoh_coh} is a sum of the incoherent and coherent contributions.

\subsection{Stationary fields}\label{Stat_Field}

\begin{figure}[tp]
\centering
\includegraphics*[align=c,width=0.3\linewidth]{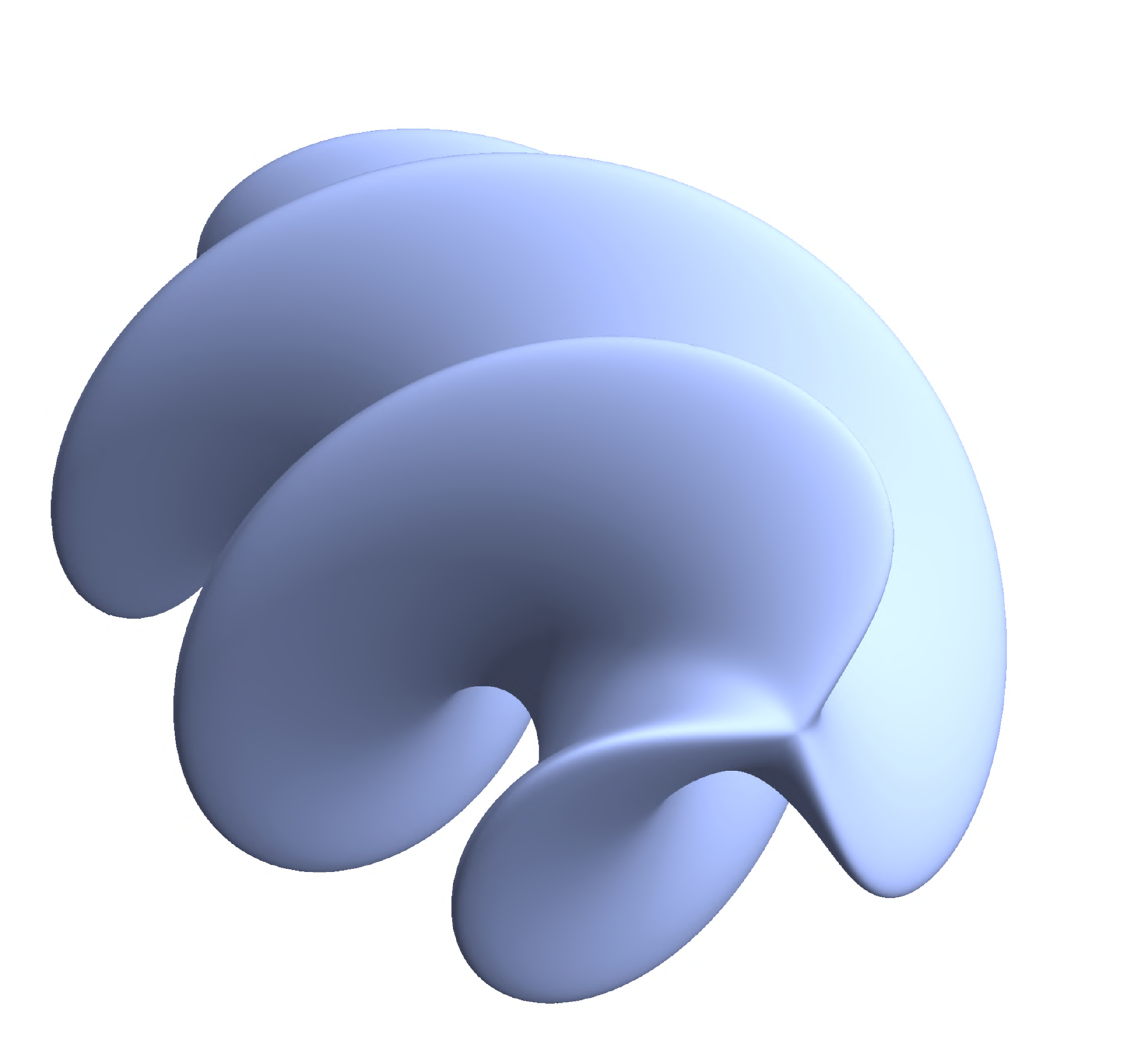}\qquad\qquad\qquad
\includegraphics*[align=c,width=0.3\linewidth]{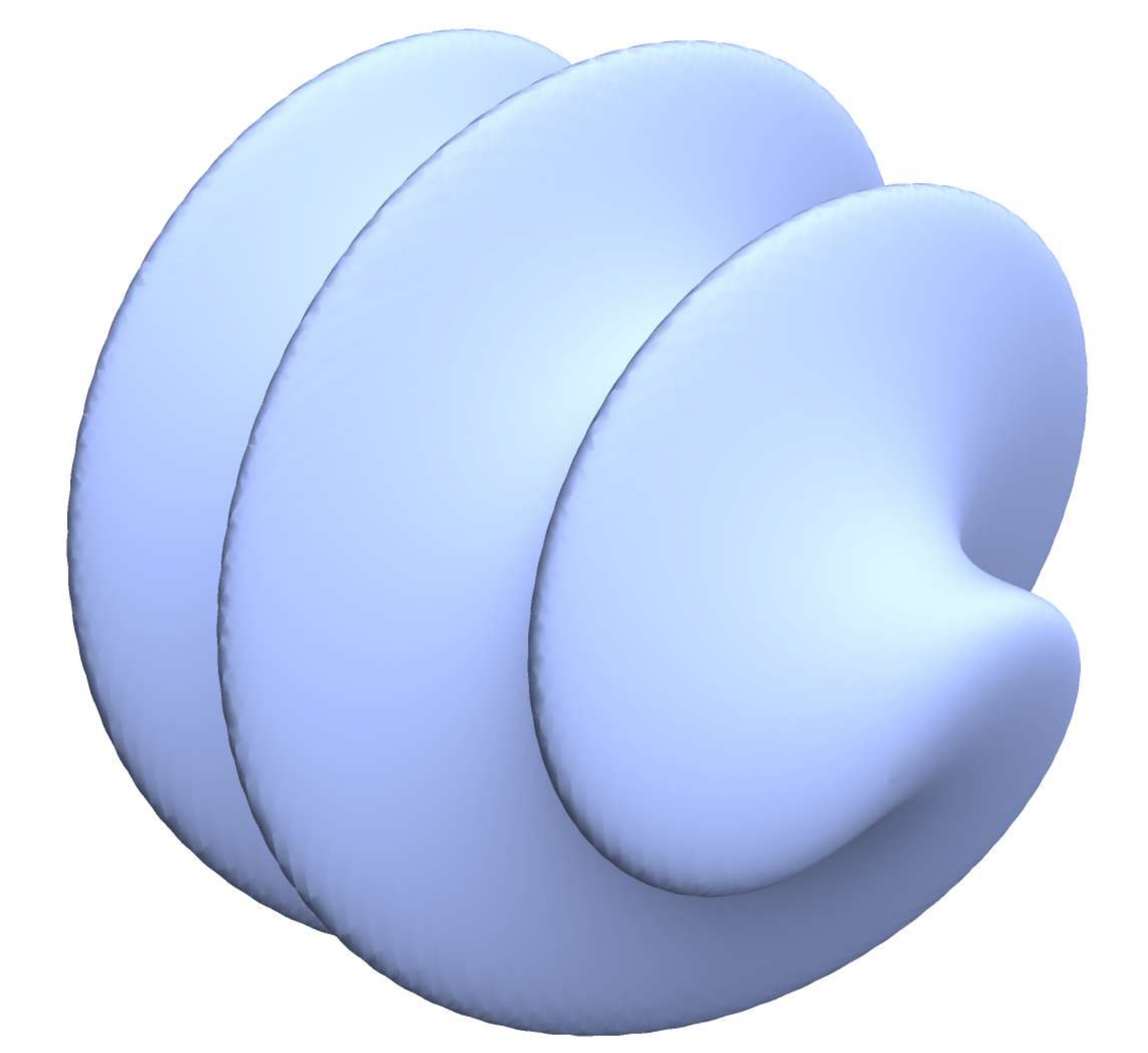}
\caption{{\footnotesize On the left panel: The surface of constant one-particle probability density \eqref{helical_bunch} with dominant harmonics $c_0$ and $c_{\pm3}$. On the right panel: The surface of constant one-particle probability density \eqref{helical_bunch} with the parameters used in Fig. \ref{CO2_ovrtk_hel_fig}.}}
\label{beam_helix_fig1}
\end{figure}

Let us consider, at first, the case of helical bunches moving in the field invariant with respect to translations \eqref{transl_statio1}. The coherent interference factor \eqref{interfer_factor} for the forward radiation is given by
\begin{equation}\label{coh_inter_fact2}
    \vf_m=\tilde{F}(k_0/\be_3-2\pi\chi m/\de)\bar{\vf}_m(k_\perp\s_\perp),
\end{equation}
where $\bar{\vf}_m(y)$ is defined in \eqref{coh_inter_red}. The amplitude of the coherent radiation is written as (see \eqref{prob_rad_incoh_coh})
\begin{equation}
    A_\rho(s,m,k_3,k_\perp)=\sum_{j=-\infty}^\infty\tilde{F}\big(k_0/\be_3-2\pi\chi (m-j)/\de\big)\bar{\vf}_{m-j}(k_\perp\s_\perp)A(0;s,l,k_3,k_\perp).
\end{equation}
If in addition to the zeroth harmonic of the Fourier expansion \eqref{helical_bunch}, which is always present, the harmonics with the numbers $\pm k$ dominate such that
\begin{equation}
    c_n\approx0,\quad n\neq\pm k, n\neq0,
\end{equation}
then the lines of the spectrum over $m$ of the coherent radiation produced by such a bunch are split up as compared to the one particle radiation spectrum. For example, if one launches such a bunch into a right-handed helical wiggler, where the one-particle radiation at the $n$-th harmonic obeys the selection rule $m=n$, then, at this harmonic, the coherent radiation of twisted photons with $m=\{n-k,n,n+k\}$ will be observed. The probability density $\rho_0(\spb)$ with dominant harmonics $\pm k$ looks as a helix with $k$ branches (see Fig. \ref{beam_helix_fig1}).

In general, one can formulate the following \emph{addition rule}: the harmonic $c_k$ of the Fourier series \eqref{helical_bunch} gives the contribution to the coherent radiation of twisted photons only with $m=j+k$, where $j$ is the projection of the total angular momentum of a twisted photon radiated by one particle moving along the center of the bunch. This rule holds for an arbitrary cold relativistic particle bunch (see Sec. \ref{Generali}), not just for helical bunches \eqref{helical_bunch}.

If the distribution $F(b_3)$ is sufficiently wide, viz.,
\begin{equation}\label{longitud_coh}
    \pi\s_3/\de\gtrsim1,
\end{equation}
where $\s_3$ is a characteristic width of the distribution $F(b_3)$, then the coherent radiation of twisted photons is concentrated near the harmonics
\begin{equation}\label{harmonics_bunch}
    k_0=2\pi\chi n \be_3/\de,\qquad \chi n>0,\;n\in \mathbb{Z},
\end{equation}
with the halfwidth
\begin{equation}
    \De k_0\lesssim \be_3/\s_3.
\end{equation}
In fact, condition \eqref{longitud_coh} ensures that harmonics \eqref{harmonics_bunch} do not overlap. For example, for the Gaussian distribution,
\begin{equation}
    F(b_3)=\frac{e^{-b_3^2/(2\s_3^2)}}{\sqrt{2\pi}\s_3},
\end{equation}
we have
\begin{equation}
    \tilde{F}(p)=e^{-p^2\s_3^2/2}.
\end{equation}
If $\s_3$ is large, then $\tilde{F}(p)$ is concentrated in a small vicinity of the point $p=0$. At the $n$-th harmonic \eqref{harmonics_bunch} induced by the bunch profile, we obtain the coherent amplitude
\begin{equation}\label{coh_ampl_bunch}
    A_\rho(s,m,k_3,k_\perp)=\bar{\vf}_{n}(k_\perp\s_\perp)A(0;s,m-n,k_3,k_\perp).
\end{equation}
Due to condition \eqref{harmonics_bunch}, the zeroth harmonic $c_0$ of the Fourier series \eqref{helical_bunch} and the respective interference factor $\bar{\vf}_0$ do not give a considerable contribution to the amplitude of the coherent radiation. Thus we see that
for sufficiently long helical bunches the \emph{strong addition rule} is fulfilled: the spectrum of twisted photons over $m$ produced by one particle is shifted by $n$ for the coherent radiation of helical bunches, where $n$ is the signed number of the coherent harmonic.

This property can be employed to design pure sources of twisted photons. As we have already discussed in the previous section, the forward radiation of one particle that was moving rectilinearly and uniformly and then has stopped instantaneously or, vice versa, was at rest and then has increased its velocity up to some nonzero value consists of twisted photons with $m=0$ \cite{BKL3}. Therefore, performing such a motion, the helical bunch of particles radiates the twisted photons with $m=n$ at the $n$-th harmonic \eqref{harmonics_bunch}. The analogous situation takes place for any one-particle radiation produced by an induced current symmetric with respect to rotations around the detector axis. In that case, the radiation of twisted photons by one particle is concentrated near $m=0$ \cite{BKL3}. Apart from the edge radiation mentioned, such a symmetry is inherent, for example, to the current density induced by a charged particle in an isotropic media when the particle moves along the detector axis. Therefore, the forward coherent VCh and transition radiations produced by a helical bunch are concentrated at harmonics \eqref{harmonics_bunch} and consist of the twisted photons with $m=n$ at the $n$-th harmonic. The total probability to record a twisted photon was already found by us and is given by formula \eqref{prob_rad_tot_hw} with $n=0$.

\begin{figure}[tp]
\centering
\includegraphics*[align=c,width=0.49\linewidth]{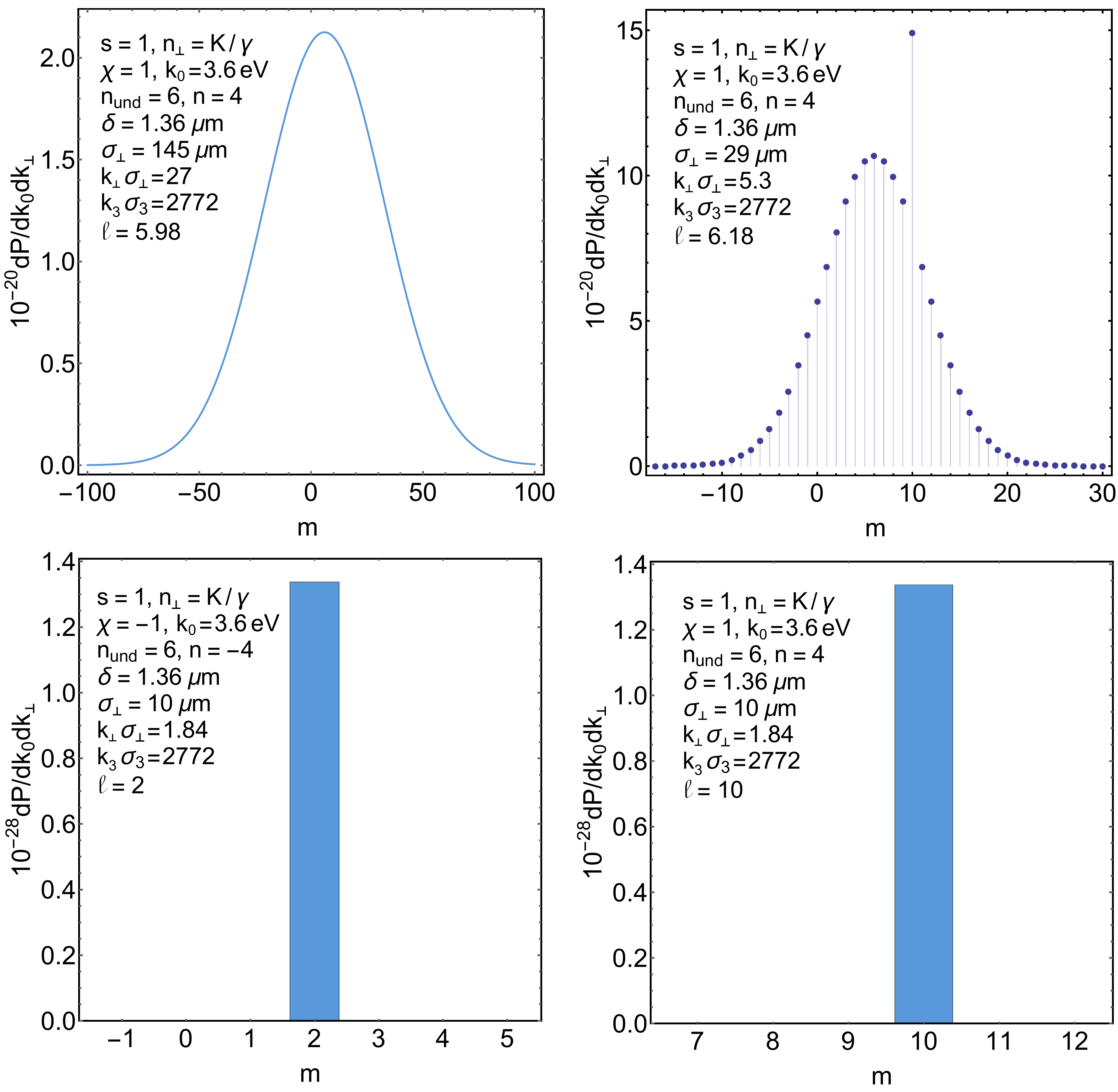}\;
\caption{{\footnotesize The radiation probability distribution over $m$ for a helical bunch of electrons moving in the helical wiggler. The Lorentz factor is $\ga=500$, the undulator strength parameter is $K=5$, and the undulator period is $\la_0=2$ cm. The parameters of the bunch are the same as in Fig. \ref{uniform bunch_fig} with the exception that now the bunch is helical \eqref{helical_bunch} and has the different waist. The dependence of the coherent contribution on the waist of the electron bunch is clearly seen. The coherent radiation of twisted photons obeys the strong addition rule and the sum rule \eqref{ang_mom_per_phot} for any photon helicity $s$. Notice that for $n_\perp<K/\gamma$ the radiation of twisted photons with $s=1$ dominates, while for $n_\perp>K/\gamma$ the twisted photons with $s=-1$ prevail \cite{BKL4}. For $n_\perp\ll1$, the orbital angular momentum $l=m-s$ can be introduced. Therefore, the peak $m=10$, for example, corresponds to twisted photons with $l=9$ for $n_\perp<K/\ga$ and to twisted photons with $l=11$ for $n_\perp>K/\ga$.}}
\label{undul_hel_helix_fig}
\end{figure}

\begin{figure}[tp]
\centering
\includegraphics*[align=c,width=0.49\linewidth]{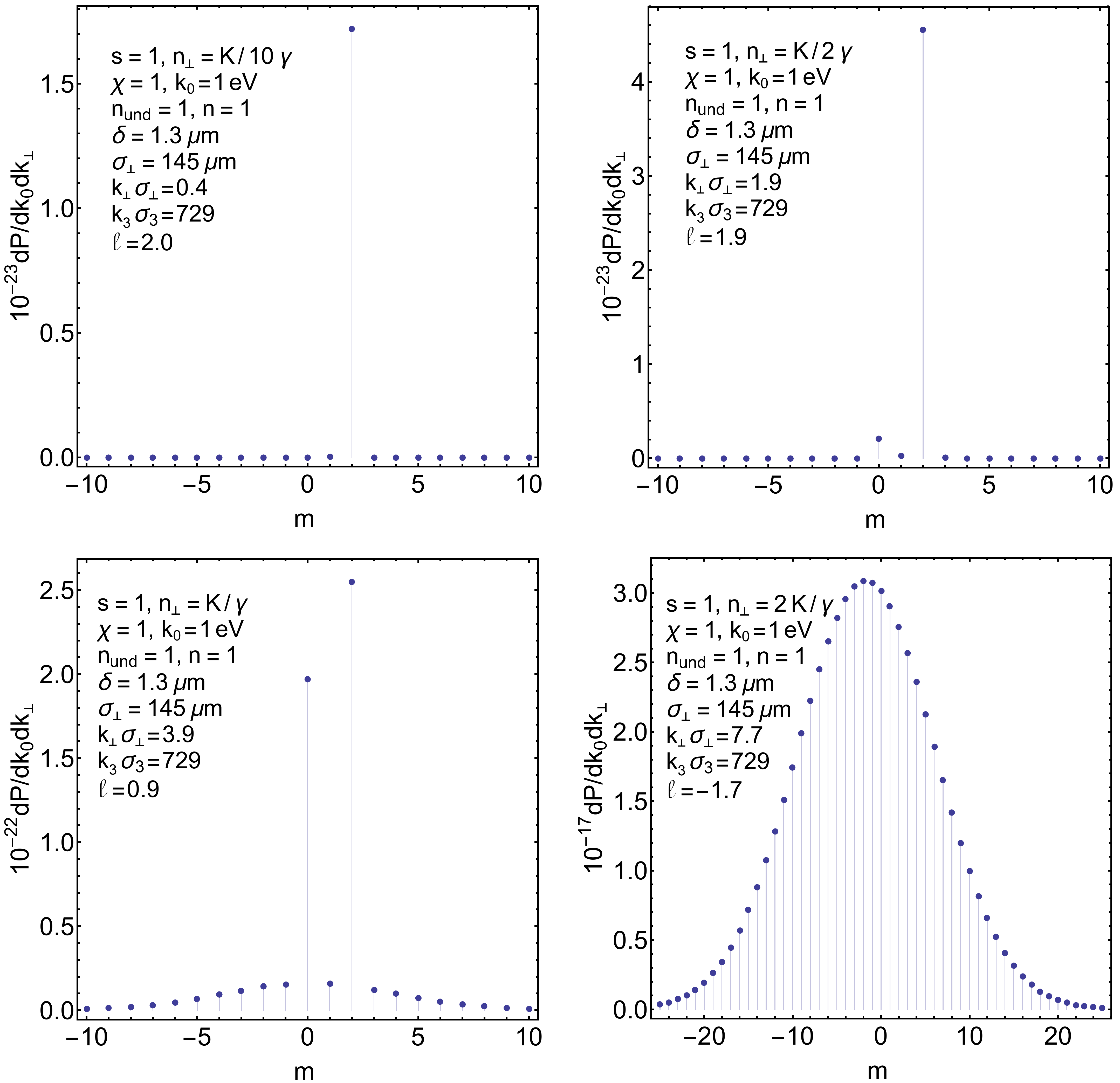}\;
\includegraphics*[align=c,width=0.49\linewidth]{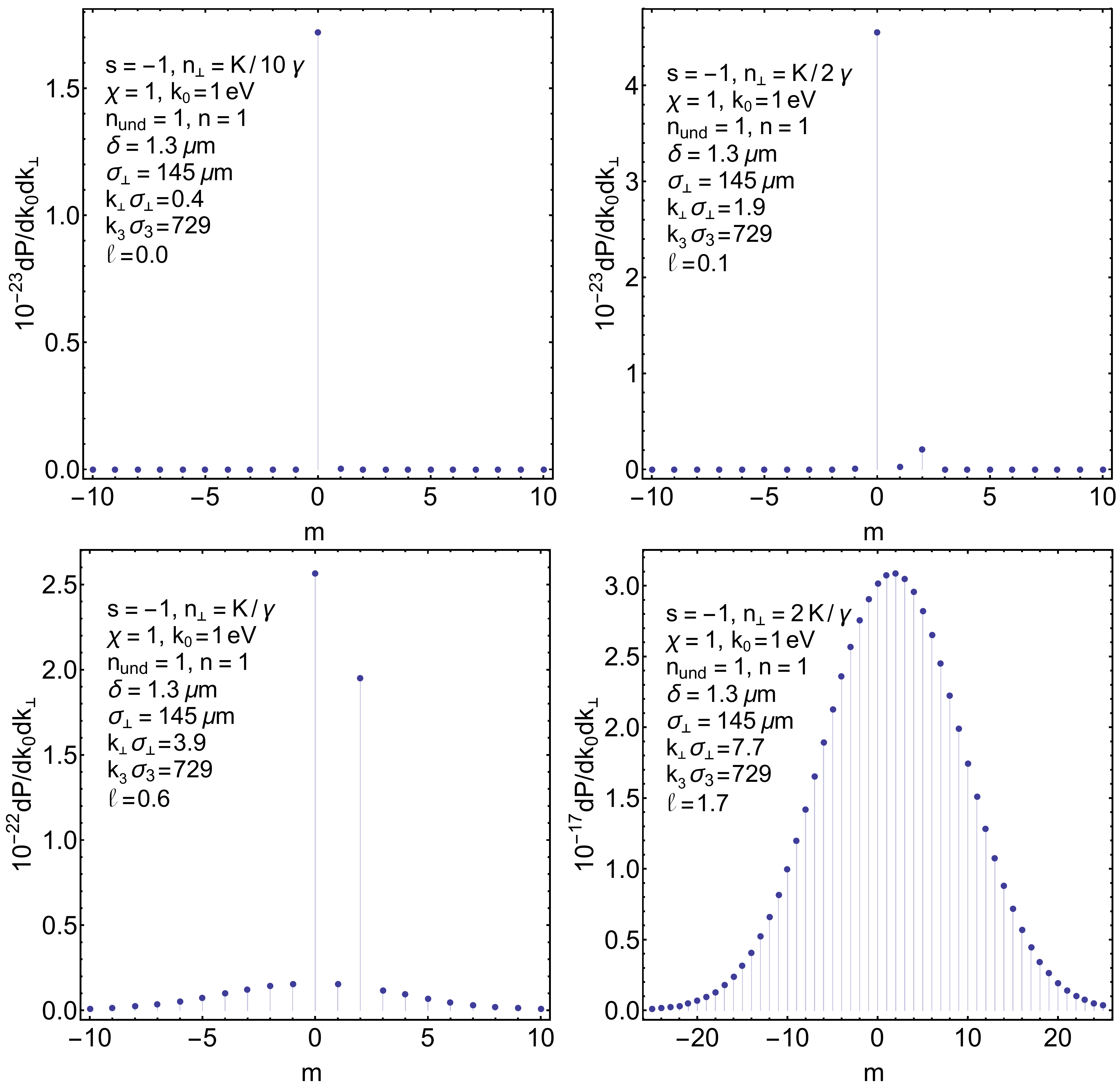}
\caption{{\footnotesize The radiation probability distribution over $m$ for a helical bunch of electrons moving in the planar undulator. The Lorentz factor $\ga=235$, the undulator strength parameter $K=1.29$, and the undulator period is $\la_0=3.3$ cm. The parameters of the electron bunch are the same as in Fig. \ref{uniform bunch_fig} with the exception that the bunch is helical \eqref{helical_bunch} and possesses the different waist. The radiation at the first undulator harmonic is considered. We see that for $n_\perp\leqslant K/\ga$ the coherent radiation dominates. This radiation obeys the strong addition rule and the sum rule \eqref{ang_mom_per_phot}. For such $n_\perp$, the orbital angular momentum $l=m-s$ can be introduced. Thus, the twisted photons with $l=\{-1,1,3\}$ are present in the radiation summed over helicities $s$. However, for $n_\perp\leqslant K/(2\ga)$, the twisted photons with $l=1$ dominate. This explains why in \cite{HKDXMHR} a clear interference pattern was observed only in a small vicinity of the origin.}}
\label{undul_fig}
\end{figure}

Notice that the VCh radiation generated by twisted electrons and incoherent Gaussian bunches of them was studied in \cite{IvSerZay,Kaminer16}. The property we have discussed above is valid for the coherent radiation of twisted photons produced by the usual plane-wave charged particles. As far as transition radiation is concerned, the generation of twisted electromagnetic waves was discussed in \cite{HemsingTR12,HemRos09} for the radiation of a helical bunch striking a metal foil. There was mentioned, in particular, the analog of the addition rule for this radiation. Nevertheless, we should stress that the both addition rules we have deduced are formulated for the projection of the total angular momentum and not for the orbital angular momentum of radiated photons. For example, the transition radiation at the first coherent harmonic \eqref{harmonics_bunch} for $\chi=1$ consists of twisted photons with $m=1$. Even in the paraxial approximation, $n_\perp\ll1$, when the orbital angular momentum can be introduced as $l=m-s$, the twisted photons with $m=1$ do not possess the orbital angular momentum $l=1$. Therefore, in contrast to \cite{HemsingTR12,HemRos09}, transition radiation at the first harmonic \eqref{harmonics_bunch} does not consist of photons with $l=1$. The summation over helicities $s$ results in the mixture of photons with $l=0$ and $l=2$. In a certain sense, one may say that the spectral line $l=1$ is split into two $l=\{0,2\}$ due to the spin of a photon. The same observations apply to the radiation produced by helical bunches in undulators discussed below (see also Figs. \ref{undul_hel_helix_fig}, \ref{undul_fig}).

In a general case, for wide distributions $F(b_3)$ satisfying \eqref{longitud_coh}, the incoherent contribution to the radiation probability is also simplified. As follows from \eqref{rho_b}, the Fourier harmonics that enter into $f_{m,m-k}$ are strongly suppressed for $k\neq0$. Therefore, the incoherent contribution to the radiation probability is almost the same as for round bunches of particles. Using relation \eqref{incoh_int_fact_round}, we find that the total probability to record a twisted photon at the $n$-th harmonic \eqref{harmonics_bunch} is
\begin{equation}\label{tot_prob}
    dP_\rho(s,m,k_3,k_\perp)\approx N\sum_{j=-\infty}^\infty f_{m-j}dP_1(s,j,k_3,k_\perp)+N(N-1)|\bar{\vf}_{n}|^2 dP_1(s,m-n,k_3,k_\perp),
\end{equation}
where $dP_1$ is the corresponding probability of radiation by one particle moving along the bunch center. Then
\begin{equation}\label{sum_rule1_1}
    \sum_{m=-\infty}^\infty dP_\rho(s,m,k_3,k_\perp)=N\big[1+(N-1)|\bar{\vf}_n|^2\big]\sum_{m=-\infty}^\infty dP_1(s,m,k_3,k_\perp),
\end{equation}
and
\begin{equation}\label{ang_mom_per_phot}
    \ell_\rho=\ell_1+n\frac{(N-1)|\bar{\vf}_n|^2}{1+(N-1)|\bar{\vf}_n|^2},
\end{equation}
where $\ell_\rho$ and $\ell_1$ are the projections of the total angular momentum per photon for the radiation produced by a helical bunch of particles and by one particle, respectively. When the coherent contribution dominates, we obviously have
\begin{equation}
    \ell_\rho\approx\ell_1+n,
\end{equation}
in accordance with the strong addition rule. When the coherent contribution is suppressed, we obtain $\ell_\rho\approx\ell_1$, in agreement with the sum rule \eqref{ang_mom_sr}.

In order to observe an intense forward coherent radiation generated by a helical bunch in undulators, it is necessary that harmonics \eqref{harmonics_bunch} overlap with the corresponding harmonics of the undulator radiation \cite{BKL2,BKL4,BaKaStrbook}
\begin{equation}\label{spectrum_sol_app}
    k_0^n=\frac{n_{und}\omega}{1-n_3\ups_3+n_{und}\omega/P_0}\equiv\frac{\bar{k}_0^n}{1+\bar{k}_0^n/P_0}\;\Leftrightarrow\; \frac{1}{k_0^n}=\frac{1}{\bar{k}_0^n}+\frac{1}{P_0},
\end{equation}
where $n_{und}$ is the harmonic number of the undulator radiation, $P_0=m\gamma$, and $\bar{k}_0^n$ is the energy of $n_{und}$-th harmonic of the undulator radiation without quantum recoil. Since $\be_3\approx1$, the helix pitch $\de$ should be a multiple of the radiation wavelength. For example, in the dipole regime, the $n$-th harmonic \eqref{harmonics_bunch} of the coherent radiation of a helical bunch has to coincide with $n_{und}$-th harmonic of the undulator radiation
\begin{equation}
    \frac{2\pi}{\de}\chi n=\frac{2\omega\ga^2n_{und}}{1+n_\perp^2\ga^2},\qquad n_\perp\ga\lesssim1.
\end{equation}
For simplicity, we neglected the quantum recoil in this formula. According to the strong addition rule, the radiation at this harmonic consists of twisted photons with $m=\{n-1,n+1\}$ (see Fig. \ref{undul_fig}). This shift of distribution over $m$ was used in \cite{HKDXMHR} to produce twisted photons in the planar undulator at the first harmonic. If the undulator is helical, then the strong addition rule entails that the radiation consists of twisted photons with $m=n\pm1$, where the sign is determined by handedness of the helix along which the electron is moving in the undulator. As for helical wigglers, we obtain the condition
\begin{equation}
    \frac{2\pi}{\de}\chi n=\frac{2\omega\ga^2n_{und}}{1+K^2+n_\perp^2\ga^2},\qquad n_\perp\ga\sim K,
\end{equation}
where $K$ is the undulator strength parameter. The coherent radiation of a helical bunch in the helical wiggler at the $n$-th  harmonic \eqref{harmonics_bunch} consists of twisted photons with $m=n\pm n_{und}$, where the sign depends on the wiggler helicity (see Fig. \ref{undul_hel_helix_fig}). The total radiation probability is given by formula \eqref{prob_rad_tot_hw}.

% отметить про выделенный случай спирального ондулятора при n_\perp<1/\gamma. в этом случае доминирует s=1 и m=1=1-l =>l=0. s=1,m=2 => l=1.
% для плоского ондулятора после сдвига l=-1,1,1,3. при n_\perp<1/\gamma вклад l=1 доминирует (он еще удваивается за счёт разных спиральностей).

Another one constraint on the bunch parameters follows from the form of the coherent interference factor $\bar{\vf}_n(k_\perp\s_\perp)$. For a small argument, we have
\begin{equation}\label{coh_fact_sml}
    \bar{\vf}_n(x)=(-1)^{(n-|n|)/2}\sum_{l=0}^\infty\frac{(-1)^l}{l!(|n|+l)!}\Big(\frac{x}{2}\Big)^{2l+|n|}M_n(2l+|n|+2)\approx (-1)^{(n-|n|)/2} \frac{M_n(|n|+2)}{|n|!}\Big(\frac{x}{2}\Big)^{|n|},
\end{equation}
i.e., the contribution of the $n$-th harmonic \eqref{harmonics_bunch} to the radiation amplitude is suppressed at small $x$. For large $x$, the coherent interference factor also tends rapidly to zero: if $\rho(\spb)$ is an infinitely differentiable function, then $\bar{\vf}_n(x)$ tends to zero faster than any power of $x^{-1}$ as $x\rightarrow\infty$. Hence, in order to generate a considerable radiation of twisted photons, one needs to impose the condition
\begin{equation}\label{sigma_perp_cond}
    \s_\perp^c\lesssim \s_\perp\lesssim \max(1,x_{max}(n))\s_\perp^c,\qquad k_\perp\s_\perp^c:=1,
\end{equation}
where $x_{max}(n)$ is the value of $x$ where $|\bar{\vf}_n(x)|$ reaches its maximum. The estimates of $x_{max}(n)$ for particular bunch profiles are given below. It turns out that $x_{max}(n)$ depends severely on the transverse profile of the bunch: the faster $\rho(\spb)$ drops to zero as $|\spb_\perp|\rightarrow\infty$, the faster $x_{max}(n)$ grows as $n$ increases. For small $n$, $x_{max}(n)$ is of order of unity. If $\s_\perp\ll \s_\perp^c$, then the probability of coherent radiation of twisted photons by a helical bunch at the $n$-th harmonic is suppressed by the factor $(\s_\perp/\s_\perp^c)^{|2n|}$. For small $|n|$, this suppression is not very strong and only the fulfillment of the second inequality in \eqref{sigma_perp_cond} is relevant.

\subsection{Electromagnetic wave}

Now we turn to the case of the forward coherent radiation created by helical bunches \eqref{helical_bunch} in the field invariant under translations \eqref{transl_wave_1} with $\be_\perp=0$ and $\theta=\{0,\pi\}$. The coherent interference factor reads
\begin{equation}
    \vf_m=\tilde{F}\Big(-\zeta k_0\frac{1-\zeta n_3}{1-\zeta\be_3}-\chi m\frac{2\pi}{\de}\Big)\bar{\vf}_m(k_\perp\s_\perp),
\end{equation}
where $\zeta:=\cos\theta=\pm1$. Then the amplitude of the coherent radiation becomes
\begin{equation}
    A_\rho(s,m,k_3,k_\perp)=\sum_{j=-\infty}^\infty\tilde{F}\Big(-\zeta k_0\frac{1-\zeta n_3}{1-\zeta\be_3}-\chi (m-j)\frac{2\pi}{\de}\Big)\bar{\vf}_{m-j}(k_\perp\s_\perp)A(0;s,j,k_3,k_\perp).
\end{equation}
All what was said about the radiation of twisted photons by helical bunches in the fields invariable under translations \eqref{transl_statio1} is fully applicable to the case at issue. In particular, the both addition rules mentioned above are satisfied. The only difference consists in the spectrum of the coherent radiation of the bunch, which now looks as
\begin{equation}\label{harmonics_bunch_ew}
    k_0=-\zeta\chi n\frac{2\pi}{\de}\frac{1-\zeta\be_3}{1-\zeta n_3},\qquad \zeta\chi n<0, n\in \mathbb{Z},
\end{equation}
where $n$ is the signed number of the coherent harmonic. The halfwidth of spectral lines is of order
\begin{equation}
    \De k_0\lesssim \s_3^{-1}\frac{1-\zeta\be_3}{1-\zeta n_3},
\end{equation}
and condition \eqref{longitud_coh} has to be met. At the $n$-th harmonic \eqref{harmonics_bunch_ew}, the amplitude of coherent radiation of a twisted photon has the form \eqref{coh_ampl_bunch}. The zeroth harmonic $c_0$ of the Fourier series \eqref{helical_bunch} does not considerably contribute to the amplitude of coherent radiation. On the other hand, the incoherent interference factors $f_{m,m-k}$ are virtually zero for $k\neq0$ and are determined solely by the the zeroth harmonic $c_0$. The total probability to record a twisted photon produced by the helical bunch of particles at the $n$-th harmonic \eqref{harmonics_bunch_ew} is given by formula \eqref{tot_prob} and the projection of the total angular momentum per photon is equal to \eqref{ang_mom_per_phot}.

%CO2_hdn
\begin{figure}[tp]
\centering
\includegraphics*[align=c,width=0.49\linewidth]{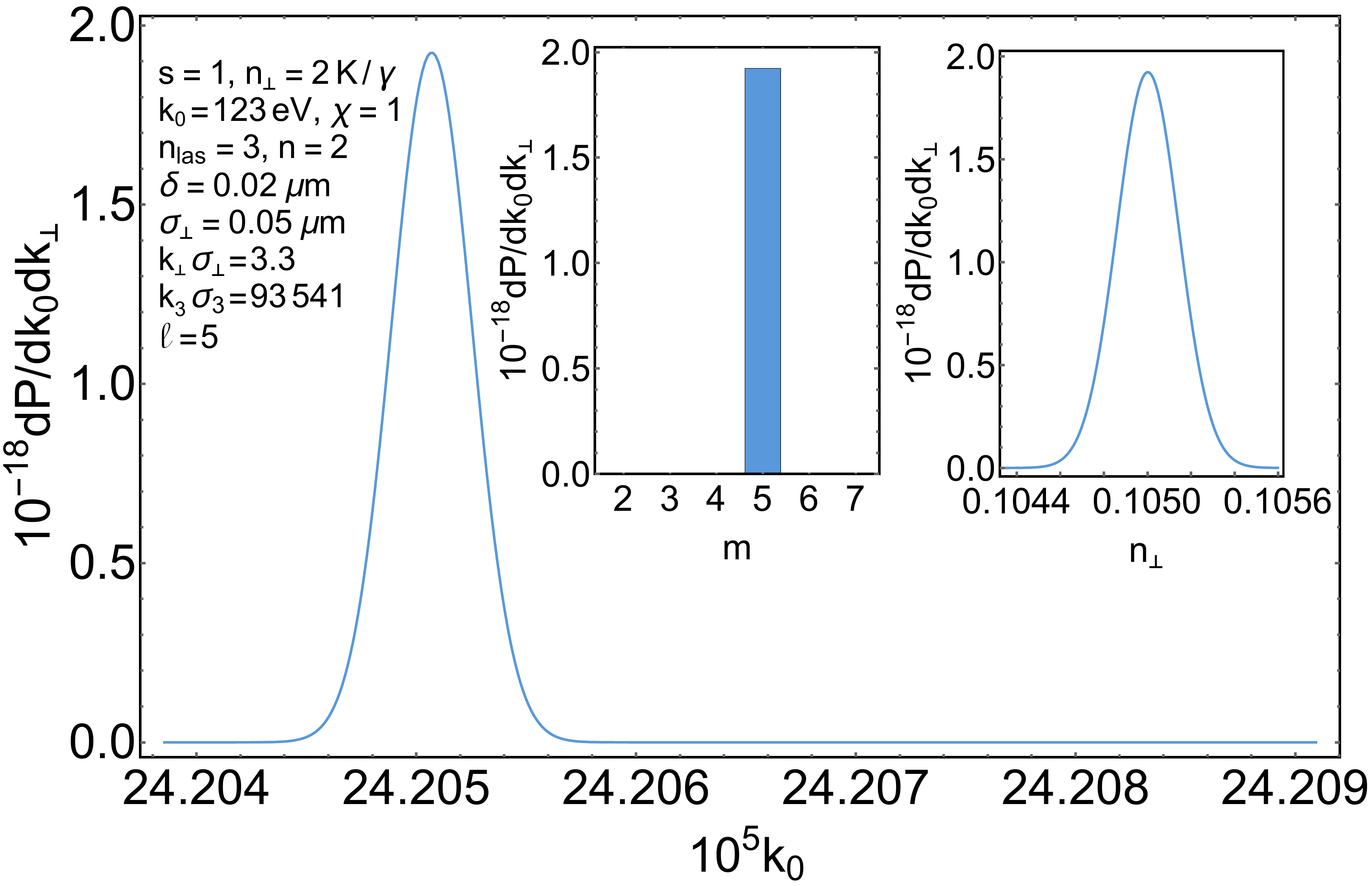}\;
\includegraphics*[align=c,width=0.49\linewidth]{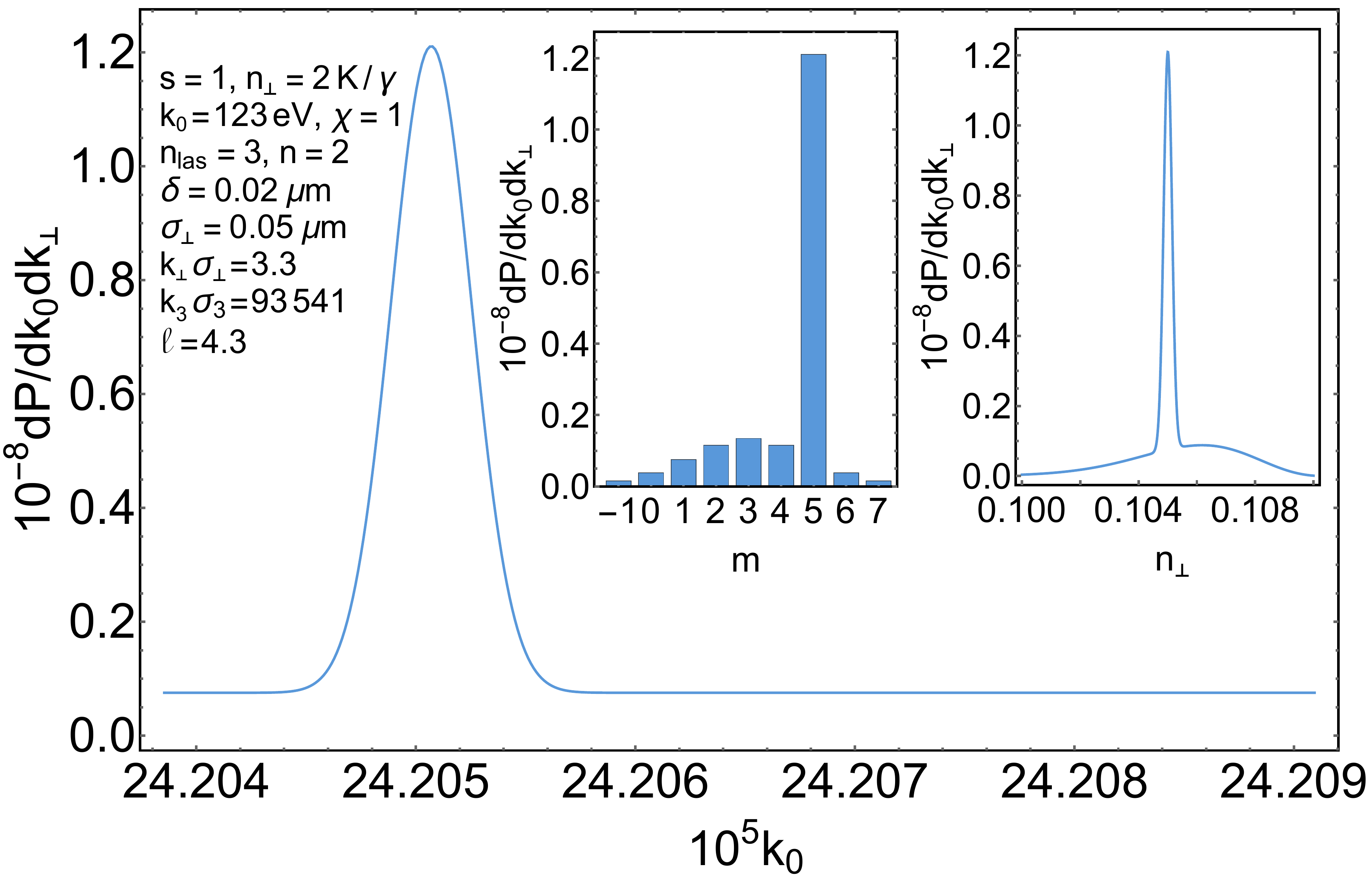}
\caption{{\footnotesize The probability of radiation of twisted photons produced by the electron bunch with $\ga=100$ in head-on collision with the circularly polarized electromagnetic wave produced by CO${}_2$ laser with the wavelength $10$ $\mu$m, the intensity $10^{18}$ W/cm${}^2$, and the amplitude envelope $f_0\sin^4(\Omega\xi/(2N))$ with $N=20$. The parameters of the electron bunch are the same as in Fig. \ref{undul_fig} with the exception that the bunch has the different waist and the number of particles. The coherent radiation dominates, the strong addition rule is satisfied, and the sum rule \eqref{ang_mom_per_phot} is fulfilled. The energy of photons is measured in the electron rest energies, $0.511$ MeV. The dependence of the radiation probability on $k_0$ and $n_\perp$ is depicted for $m$ taken at the maximum, i.e., $m=5$. On the left panel: The number of particles in the bunch is $3\times10^8$. On the right panel: The number of particles in the bunch is $200$.}}
\label{CO2_hdn_hel_fig}
\end{figure}

\begin{figure}[tp]
\centering
\includegraphics*[align=c,width=0.49\linewidth]{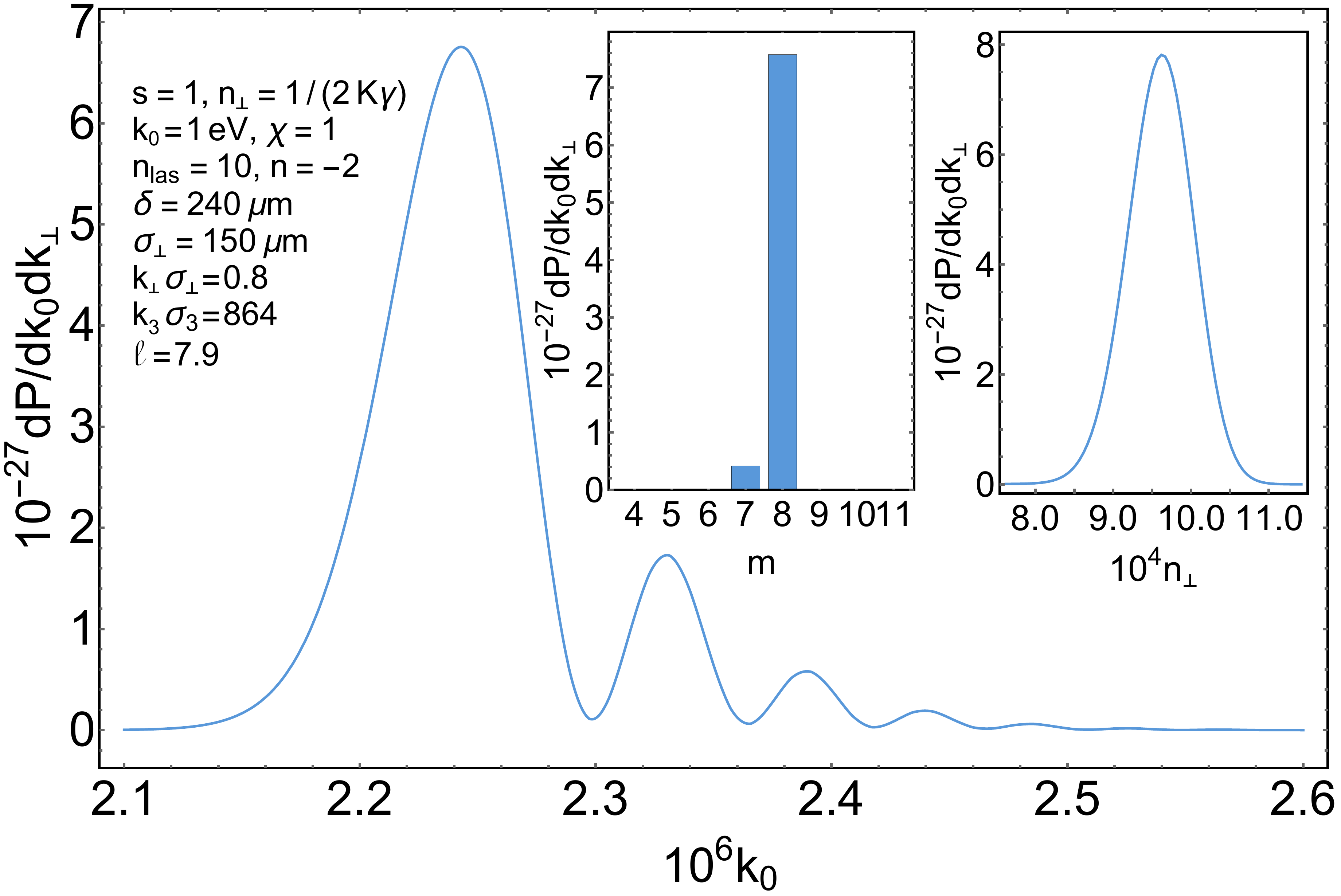}\;
\includegraphics*[align=c,width=0.49\linewidth]{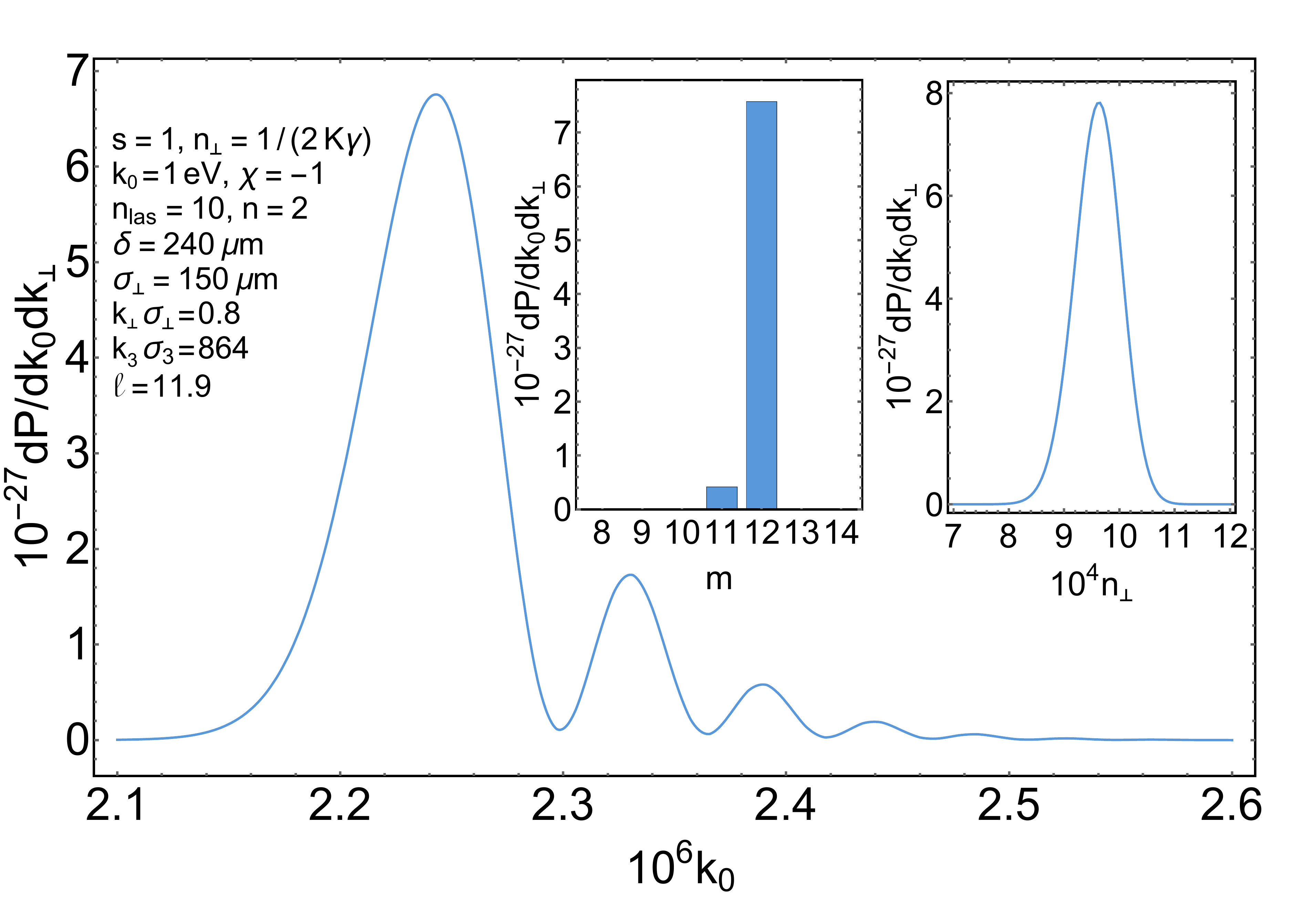}
\caption{{\footnotesize The same as in Fig. \ref{CO2_hdn_hel_fig} but in the case when the electromagnetic wave overtakes the electron bunch. The parameters of the electron bunch are the same as in Fig. \ref{CO2_hdn_hel_fig} with the exception that the bunch has the different waist and the helix pitch, and the number of particles $3\times10^8$. The surface of constant one-particle probability density is depicted in Fig. \ref{beam_helix_fig1}. The coherent radiation dominates, the strong addition rule is satisfied, and the sum rule \eqref{ang_mom_per_phot} is fulfilled. The energy of photons is measured in the electron rest energies, $0.511$ MeV. The length of the interaction region is $L=3$ m. The dependence of the radiation probability on $k_0$ and $n_\perp$ is depicted for $m$ taken at the maximum, i.e., $m=8$ for the left panel and $m=12$ for the right panel.}}
\label{CO2_ovrtk_hel_fig}
\end{figure}

In particular, the strong addition rule for radiation of helical bunches in the laser wave with circular polarization says that the projection of the total angular momentum of a twisted photon radiated at the $n$-th harmonic \eqref{harmonics_bunch_ew} is $m=n\pm n_{las}$, where $n_{las}$ is the harmonic number of radiation produced by one particle in the laser wave and the sign in front of $n_{las}$ is determined by the laser wave helicity (see Figs. \ref{CO2_hdn_hel_fig}, \ref{CO2_ovrtk_hel_fig}). In this case, the total probability to record a twisted photon is described by formula \eqref{prob_rad_tot_hw}.

For $\zeta=1$, in the case of forward radiation in a laser wave with smooth envelope, the radiation spectrum has the form \cite{BKL4}
\begin{equation}\label{spectrum_forw_1}
    k_0\approx \Omega n_{las}\Big[1+\frac{n_\perp^2}{4\ups_-^2}(1+K^2)\Big]^{-1},\qquad n_{las}=\overline{1,\infty},
\end{equation}
where $\ups_-:=\ga(1-\be_3)=const$ and $K:=f_0/(m_e\Omega)$ is the undulator strength parameter. Here $f_0$ is the amplitude of the laser wave strength field, $m_e$ is the mass of a charged particle, and $\Omega$ is the energy of photons in this wave. Then, in the ultrarelativistic regime, the harmonics of coherent radiation \eqref{harmonics_bunch_ew} coincide with \eqref{spectrum_forw_1} when
\begin{equation}
    -\frac{2\pi}{\de}\frac{\chi n}{n_\perp^2\ga^2}\approx \frac{|\Omega|n_{las}}{1+n_\perp^2\ga^2(1+K^2)}.
\end{equation}
The constraints on $\s_\perp$ are obtained by substitution of \eqref{spectrum_forw_1} into \eqref{sigma_perp_cond}. In particular, in the wiggler case, $K\gtrsim3$, for the optimum value of $n_\perp$ \cite{BKL4},
\begin{equation}\label{n_perp_opt_1}
    n_\perp=2\ups_-/K\approx 1/(\ga K),
\end{equation}
we come to
\begin{equation}
    -\chi n K^2\frac{2\pi}{\de}\approx \frac{|\Omega|}{2}n_{las},\qquad\s^c_\perp\sim \ga K/k_0.
\end{equation}
As is seen, $\de$ should be of order of the radiation wavelength times $K^2$. The optimum value of $\s_\perp$ has to be approximately  the radiation wavelength times $\ga K$. Besides, the length of the interaction region is of order
\begin{equation}
    L=\s_3/(1-\be_3)\approx 2\s_3\ga^2.
\end{equation}
It, of course, should be much smaller than the corresponding size of the experimental facility. The plots of the radiation probability in this case are presented in Fig. \ref{CO2_ovrtk_hel_fig}.

%CO2_ovrtk

For $\zeta=-1$, the radiation spectrum of the forward radiation is given by \cite{TaHaKa,KatohPRL,TairKato,BKL4}
\begin{equation}\label{spectrum_forw_-1}
    k_0\approx\frac{\Omega n_{las}\ups_-^2}{1+K^2+n^2_\perp\ups_-^2/4},\qquad n_{las}=\overline{1,\infty},
\end{equation}
where $\ups_-:=\ga(1+\be_3)=const$. As a result, we have the condition
\begin{equation}
    \frac{2\pi}{\de}\chi n\approx\frac{4|\Omega|\ga^2n_{las}}{1+K^2+n_\perp^2\ga^2}.
\end{equation}
Hence, $\de$ should be of order of the radiation wavelength. The optimum value of $\s_\perp$ is found from \eqref{sigma_perp_cond}. In the wiggler regime, $K\gtrsim3$, $n_\perp\ga\approx K$, and we obtain
\begin{equation}
    \frac{2\pi}{\de}\chi n\approx\frac{2|\Omega|\ga^2}{K^2}n_{las},\qquad\sigma^c_\perp\sim\frac{\ga}{k_0 K}.
\end{equation}
Thus, in the wiggler regime, the optimum value of $\s_\perp$ is of order of the radiation wavelength times $\ga/K$. For $\s_\perp\ll\s_\perp^c$, the radiation probability of twisted photons is suppressed by the factor $(\s_\perp/\s_\perp^c)^{|2n|}$. For small $|n|$, this suppression is rather weak and the waist of the particle bunch is constrained only by the second inequality in \eqref{sigma_perp_cond}. The plots of the radiation probability in this case are given in Fig. \ref{CO2_hdn_hel_fig}.

\subsection{Explicit expressions}\label{Expl_Expr_Coh}

\begin{figure}[tp]
\centering
\includegraphics*[align=c,width=0.9\linewidth]{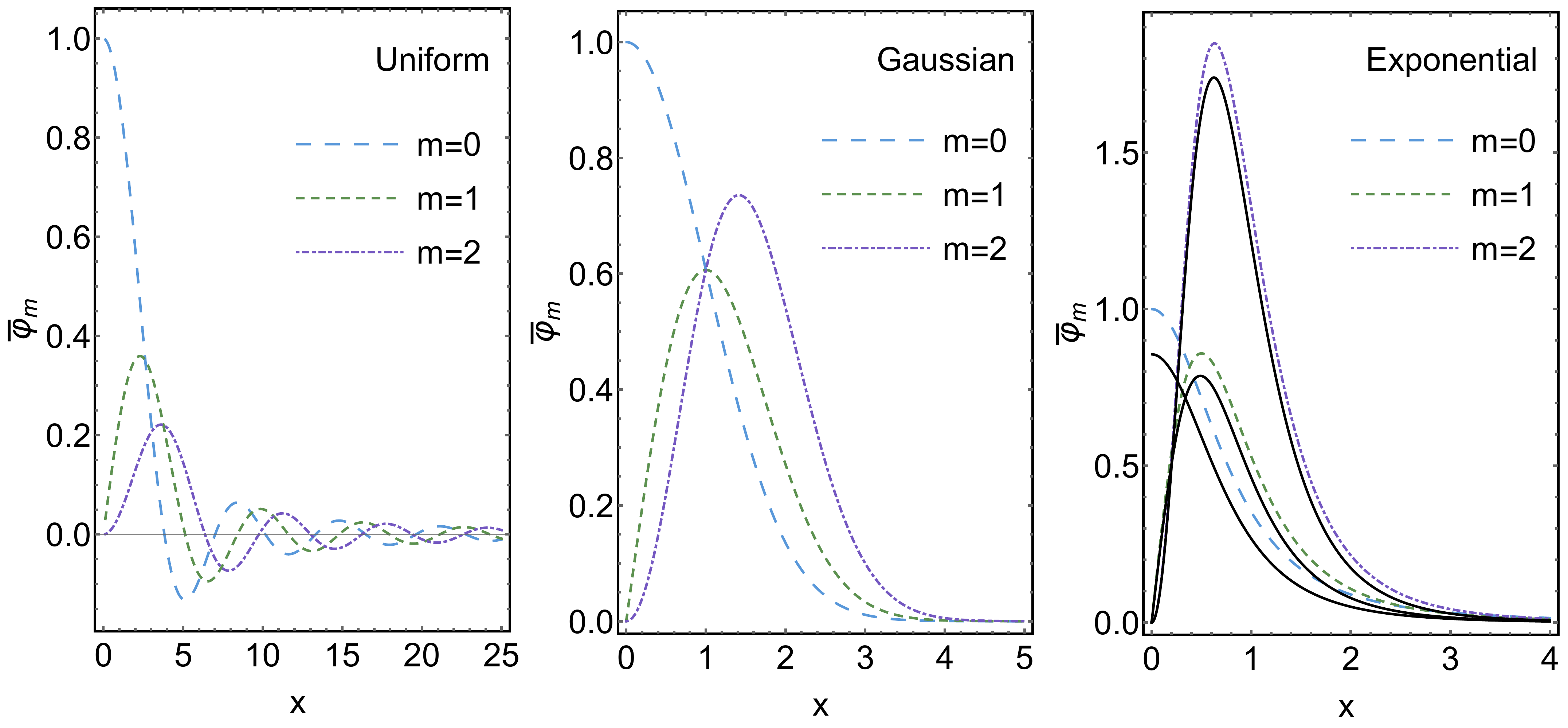}
\caption{{\footnotesize The normalized coherent interference factor $\bar{\vf}_m(x)/\al_m$ for different bunch profiles. The thin black lines on the right plot depict the normalized coherent interference factor for the generalized exponential profile with $\la=2/3$.}}
\label{form_factor_coh_fig}
\end{figure}

Let us find the explicit expressions for coherent interference factors $\bar{\vf}_m(x)$ for the simple radial profiles $c_m(r)$ considered in the previous section.

a) Uniform distribution \eqref{uniform_distr}:
\begin{equation}\label{coh_fact_uni}
    \bar{\vf}_m(x)=\al_m(-1)^{(m-|m|)/2}\frac{2J_{|m|+1}(x)}{x},
\end{equation}
where $x=k_\perp\s_\perp$. For $x=\ga_{|m|+1,k}$, $k=\overline{1,\infty}$, where $\ga_{|m|+1,k}$ are zeros of the Bessel function, the coherent interference factor vanishes, i.e., the radiation of twisted photons with such  $k_\perp\s_\perp$ is absent. For small $x$, it follows from general formula \eqref{coh_fact_sml} that
\begin{equation}
    \bar{\vf}_m(x)\approx\al_m\frac{(-1)^{(m-|m|)/2}}{|m+1|!}\Big(\frac{x}{2}\Big)^{|m|}+\cdots.
\end{equation}
When $x\gg\max(1,|m|)$, the coherent interference factor behaves as  $\bar{\vf}_m(x)\sim x^{-3/2}$ and the respective contribution to the radiation probability decreases as $x^{-3}$. Since the distribution $\rho(\spb)$ corresponding to \eqref{uniform_distr} is not smooth at $|\spb_\perp|=\s_\perp$,  $\bar{\vf}_m(x)$ tends to zero at large argument only as a power. Nevertheless, its contribution to the radiation probability decreases faster than the incoherent contribution does for large $x$. This example shows that sharp edges of the probability distribution $\rho(\spb)$ improves the coherent properties of hard photon radiation. The plots of $\bar{\vf}_m(x)$ for different $m$ are presented in Fig. \ref{form_factor_coh_fig}.

For large $|m|$, the maximum of $|\bar{\vf}_m(x)|$ is located at
\begin{equation}
    x_{max}(m)\approx\ga'_{|m|+1,1}-0.62(|m|+1)^{-1/3}+\cdots,\qquad\ga'_{|m|+1,1}\approx|m|+1+0.81(|m|+1)^{1/3}+\cdots,
\end{equation}
where $\ga'_{|m|+1,1}$ is the first positive zero of $J'_{|m|+1}(x)$. Some properties of the extrema of \eqref{coh_fact_uni} can be found in \cite{LandBess}. The maximum value \cite{NIST},
\begin{equation}
    |\bar{\vf}_m(x_{max}(m))|/|\al_m|\approx 1.35|m+1|^{-4/3},
\end{equation}
tends rather fast to zero at large $|m|$.

b) Gaussian bunch \eqref{Gauss_distr}:
\begin{equation}
    \bar{\vf}_m(x)=\al_m(-1)^{(m-|m|)/2}x^{|m|}e^{-x^2/2}.
\end{equation}
For small $x$, we obtain
\begin{equation}
    \bar{\vf}_m(x)=\al_m(-1)^{(m-|m|)/2} x^{|m|}+\cdots.
\end{equation}
For $x\gg\max(1,\sqrt{|m|})$, the coherent interference factor tends rapidly to zero. The maximum of $|\bar{\vf}_m(x)|$ is located at
\begin{equation}
    x_{max}(m)=|m|^{1/2}.
\end{equation}
The maximum value,
\begin{equation}
    |\bar{\vf}_m(x_{max}(m))|/|\al_m|=e^{|m|(\ln|m|-1)/2},
\end{equation}
grows as $|m|$ increases. The plots of $\bar{\vf}_m(x)$ are given in Fig. \ref{form_factor_coh_fig}.

c) Exponential profile \eqref{exp_distr}:
\begin{equation}\label{coh_fact_exp}
    \bar{\vf}_m(x)=\al_m(-1)^{(m-|m|)/2}\frac{2}{\sqrt{\pi}} \frac{(2x)^{|m|} \Ga(|m|+3/2)}{(1+x^2)^{|m|+3/2}}.
\end{equation}
For small $x$, we have
\begin{equation}
    \bar{\vf}_m(x)\approx\al_m(-1)^{(m-|m|)/2}\frac{2}{\sqrt{\pi}}\Ga(|m|+3/2)(2x)^{|m|}+\cdots.
\end{equation}
For $x\gg1$, the coherent interference factor behaves as
\begin{equation}
    \bar{\vf}_m(x)\approx\al_m(-1)^{(m-|m|)/2}\frac{2^{1+|m|}}{\sqrt{\pi}}\Ga(|m|+3/2)x^{-|m|-3}.
\end{equation}
The powerlike decrease law is a consequence of the fact that the first derivative of the probability density $\rho(\spb)$ corresponding to  \eqref{exp_distr} possesses a discontinuity at $\spb=0$. Nevertheless, $|\bar{\vf}_m(x)|$ quickly diminishes as $x$ increases. The function $|\bar{\vf}_m(x)|$ reaches its maximum at
\begin{equation}\label{xmax_exp}
    x_{max}(m)=\sqrt{\frac{|m|}{|m|+3}}<1.
\end{equation}
For large $|m|$, the maximum value,
\begin{equation}\label{vfmax_exp}
    |\bar{\vf}_m(x_{max}(m))|/|\al_m|\approx |m|e^{|m|(\ln\frac{|m|}{2}-1)},
\end{equation}
rapidly increases as $|m|$ increases. The plots of $\bar{\vf}_m(x)$ for different $m$ are presented in Fig. \ref{form_factor_coh_fig}.

d) As for generalized exponential profile \eqref{gen_exp_profile}, we deduce \cite{PruBryMar2}\footnote{Notice that formula [(2.12.1.6), \cite{PruBryMar2}] contains misprints (cf. [(2.12.23.2), \cite{PruBryMar2}]).}
\begin{equation}\label{gen_exp_coh}
    \bar{\vf}_m(x)=\al_m(-1)^{(m-|m|)/2}\sqrt{\frac{2}{\pi}} x^{|m|}\lambda^{|m|+3/2}(1+x^2)^{-|m|/2-3/4} K_{|m|+3/2}\big(\la\sqrt{1+x^2}\big).
\end{equation}
For $\la=0$, expression \eqref{gen_exp_coh} is reduced to \eqref{coh_fact_exp}. Notice that the Macdonald function entering into \eqref{gen_exp_coh} is expressed through elementary functions. For small $x$, we, evidently, have
\begin{equation}
    \bar{\vf}_m(x)\approx \al_m(-1)^{(m-|m|)/2}\sqrt{\frac{2}{\pi}} x^{|m|}\lambda^{|m|+3/2} K_{|m|+3/2}(\la)+\cdots.
\end{equation}
For $x\gg1$, $\la x\gg\max(1,|m|)$, we obtain
\begin{equation}
    \bar{\vf}_m(x)\approx \al_m(-1)^{(m-|m|)/2}\la^{|m|+1} e^{-\la x}/x^2.
\end{equation}
Thus, for large $x$, the coherent interference factor tends rapidly to zero. When $\nu\rightarrow+\infty$, the following asymptotic representation takes place \cite{NIST}:
\begin{equation}
    K_\nu(z)\approx\sqrt{\frac{\pi}{2\nu}}\Big(\frac{ez}{2\nu}\Big)^{-\nu}.
\end{equation}
Substituting this expression into \eqref{gen_exp_coh}, we see that expressions \eqref{coh_fact_exp} and \eqref{gen_exp_coh} coincide in the limit $|m|\rightarrow\infty$. Therefore, for large $|m|$, $|\bar{\vf}_m(x)|$ reaches its maximum at the point \eqref{xmax_exp}. The maximum value of $|\bar{\vf}_m(x)|$ for large $m$ is given by formula \eqref{vfmax_exp}. The plots of $\bar{\vf}_m(x)$ for different $m$ and $\la$ and their comparison with the coherent interference factor for the exponential profile are presented in Fig. \ref{form_factor_coh_fig}.

\subsection{Generalization}\label{Generali}

Let us consider the forward coherent radiation of a particle bunch with an arbitrary profile $\rho(\spb)$ moving in the field that is invariant under translations \eqref{transl_statio1} with $\be_\perp=0$ and $\theta=\{0,\pi\}$. Developing the one-particle probability density as a Fourier series
\begin{equation}\label{gen_profil}
    \rho(\spb)=\sum_{n=-\infty}^\infty c_n(b_\perp/\s_\perp,b_3)e^{in\psi},\qquad c^*_n(b_\perp/\s_\perp,b_3)=c_{-n}(b_\perp/\s_\perp,b_3),
\end{equation}
and substituting it into \eqref{interfer_factor}, we obtain
\begin{equation}
    \vf_m=2\pi\s_\perp^2\int_0^\infty drr\tilde{c}_m(r,k_0/\be_3)J_m(xr),
\end{equation}
where $x:=k_\perp\s_\perp$ and
\begin{equation}\label{Fourier_Fourier}
    \tilde{c}_m(r,k_0/\be_3)=\int db_3c_m(r,b_3)e^{-ik_0b_3/\be_3},\qquad \tilde{c}^*_m(r,k_0/\be_3)=\tilde{c}_{-m}(r,-k_0/\be_3).
\end{equation}
The kernel $\tilde{c}_{m-n}(r,k-p)$ is Hermitian positive-definite with respect to the variables $(m,k)$, $(n,p)$. Then the amplitude of the coherent radiation becomes
\begin{equation}
    A_\rho(s,m,k_3,k_\perp)=\sum_{j=-\infty}^\infty \vf_{m-j}(k_\perp\s_\perp)A(0;s,j,k_3,k_\perp).
\end{equation}
As we see, the forward coherent radiation of a particle bunch \eqref{gen_profil} obeys the addition rule discussed above.

In order to obtain a pure source of twisted photons, we demand that $c_k(r,p)$ are concentrated near the points $p^{(i)}_k$, $i=\overline{1,n_k}$, and the peaks of $c_k(r,p)$ do not overlap for different $k$. Due to the symmetry property \eqref{Fourier_Fourier}, the following relation holds
\begin{equation}
    p^{(i)}_k=-p^{(i)}_{-k}.
\end{equation}
Then, the addition rule leads to the strong addition rule, i.e., the coherent radiation at the photon energy
\begin{equation}
    k_0=\be_3 p^{(i)}_k,
\end{equation}
where only positive $p^{(i)}_k$ are taken, consists of twisted photons with $m=j+k$, where $j$ is the projection of the total angular momentum of twisted photons radiated by the particle moving along the center of the bunch. Performing the inverse Fourier transform of $c_k(r,p)$, it is not difficult to see that this situation happens when $\rho(\spb)$ is a superposition of helical bunches \eqref{helical_bunch} with different $\de$ and $\chi$. This source of twisted photons is pure when the one-particle amplitude is lumped near some total angular momentum projection $l=l_0$.

Similarly, the forward coherent radiation produced by a bunch of particles moving in the field invariant with respect to translations  \eqref{transl_wave_1} with $\be_\perp=0$ and $\theta=\{0,\pi\}$ can be considered. The analysis is quite analogous and we do not dwell on it here.

\section{Conclusion}

Let us sum up the main results. We derived simple formula \eqref{prob_rad_incoh_coh} for the probability to record a twisted photon produced by a cold relativistic bunch of charged particles. This formula includes both incoherent and coherent contributions and allows one to describe the radiation produced by bunches of particles using only the one-particle radiation amplitude. Of course, such a simple approach is valid only in the case when, initially, the particles in the bunch have almost the same momenta and the interaction between them can be neglected on the radiation formation scale.

Then we particularized the general formula to the two cases: the bunch of particles strikes a stationary electromagnetic field, which is invariant with respect to translations perpendicular to some distinguished vector on the scale of the bunch waist; the bunch of particles hits a propagating electromagnetic wave, the electromagnetic wave being plane in the vicinity of the bunch. Eventually, we consider only the forward radiation, when the bunch falls normally onto the constant external field surfaces and the direction of its initial motion is parallel to the axis of the twisted photon detector. Notice that, in all these cases, the parallel transport of the particle trajectory is a symmetry of the Lorentz equations with such fields.

The coherent and incoherent contributions prove to depend on the corresponding interference factors \eqref{interfer_factor}. We investigated separately the properties of these interference factors. In particular, we generalized slightly the sum rules for the incoherent radiation presented in \cite{BKb}. As an example, we considered the radiation produced by cold relativistic particle bunches in a dispersive isotropic medium (the edge, transition, and VCh radiations) and in undulators and wigglers. In particular, we found that for the edge, transition, and VCh radiations, and for the radiation by bunches of particles in the helical undulators and wigglers the incoherent contribution to the radiation of twisted photons is determined solely by the zeroth Fourier harmonic of the bunch distribution with respect to the azimuth angle, i.e., the probability of incoherent radiation is the same as for round bunches studied in \cite{BKb}. As for planar undulators and wigglers, we obtained that the odd Fourier harmonics of the bunch distribution with respect to the azimuth angle do not contribute to the incoherent radiation of twisted photons. Hence, in this case, the incoherent radiation is the same as for the bunch symmetric under the reflection with respect to the detector axis. This property is valid for arbitrary energies of twisted photons produced by planar undulators. However, at large energies, it is universal.

Namely, we obtained the general asymptotics of the incoherent interference factor in Sec. \ref{Expl_Expr_Incoh} and it turned out that, at large photon energies, the contribution of odd Fourier harmonics of the bunch distribution with respect to the azimuth angle is suppressed in comparison with the contribution of even harmonics. Therefore, at large photon energies where the incoherent contribution dominates, we have the same situation as in the case of planar undulators (see Figs. \ref{uniform bunch_fig}, \ref{uniform bunch1_fig}). The explicit expressions for the incoherent interference factor for several simple bunch profiles were also derived in Sec. \ref{Expl_Expr_Incoh} so that the general asymptotics at high energies was explicitly confirmed.

As for the coherent part of radiation, we established in Secs. \ref{Stat_Field}, \ref{Generali} the addition rule that holds for the coherent radiation by arbitrary bunches of particles moving along parallel trajectories. In general, the coherent radiation depends severely on the bunch profile. Therefore, to be more specific, we restricted our consideration to the case of helical bunches \cite{RibGauNin14,HemMar12,HemMarRos11,HemRos09,HemStuXiZh14,HKDXMHR,HemsingTR12,CLiu16,XLZhu18,LBJu16}. The addition rule applied to sufficiently long helical bunches results in the strong addition rule stating that the spectrum of twisted photons over the projection of the total angular momentum $m$ produced by one particle is shifted by $n$, where $n$ is the signed harmonic number of coherent radiation produced by the helical bunch (see Sec. \ref{Stat_Field}). Similar, but not exactly the same, observations were made in \cite{RibGauNin14,HemMar12,HemMarRos11,HemRos09,HemStuXiZh14,HKDXMHR,HemsingTR12} in studying the properties of radiation produced by helical bunches (see Figs. \ref{undul_hel_helix_fig}, \ref{undul_fig}). This property can be employed for elaboration of bright pure sources of twisted photons. We investigated several examples of radiation produced by helical bunches including the edge, transition, and VCh radiations, the helical undulator radiation, and the radiation generated by charged particles in the laser wave with circular polarization (see Figs. \ref{CO2_hdn_hel_fig}, \ref{CO2_ovrtk_hel_fig}). In these cases, we revealed the sum rules for the total probability to a record a twisted photon \eqref{sum_rule1_1} and for the projection of the total angular momentum per photon \eqref{ang_mom_per_phot}. In Sec. \ref{Expl_Expr_Coh}, we obtained the explicit expressions for the coherent interference factor for simple bunch profiles. This allowed us to find the optimum parameters of a helical bunch to generate the twisted photons.

\paragraph{Acknowledgments.}

This work is supported by the Russian Science Foundation (project No. 17-72-20013).

%\newpage\selectlanguage{english}


\begin{thebibliography}{999}

%twphts by bunches. expr techniques%EAllaria08,HKDXMHR,BHKMSS,KatohSRexp,Hemsing14hel,HemsingTR12,HemStuXiZh14,Rubic17

%beam manipul. coherent radiation. helical undulator radiator. 2nd harmonic. no twphtns
\bibitem{EAllaria08}
E. Allaria \textit{et al}.,
Experimental characterization of nonlinear harmonic generation in planar and helical undulators,
Phys. Rev. Lett. \textbf{100}, 174801 (2008).

%beam manipul. coherent radiation. helical bunches. transition radiation. beam diagnostics
\bibitem{HemsingTR12}
E. Hemsing \textit{et al}.,
Experimental observation of helical microbunching of a relativistic electron beam,
Appl. Phys. Lett. \textbf{100}, 091110 (2012).

%beam manipul. coherent radiation. helical bunches. radiating undulator (with any polarization)
\bibitem{HKDXMHR}
E. Hemsing \textit{et al}.,
Coherent optical vortices from relativistic electron beams,
Nature Phys. \textbf{9}, 549 (2013).

%2nd harmonics, usual bunch, w/o modulator
\bibitem{BHKMSS}
J. Bahrdt \textit{et al}.,
First observation of photons carrying orbital angular momentum in undulator radiation,
Phys. Rev. Lett. \textbf{111}, 034801 (2013).

%beam manipul. coherent radiation. helical undulator radiator. 2nd harmonic. sel. rule
\bibitem{Hemsing14hel}
E. Hemsing \textit{et al}.,
First characterization of coherent optical vortices from harmonic undulator radiation,
Phys. Rev. Lett. \textbf{113}, 134803 (2014).

%beam manipul. coherent radiation. helical bunches. tw phtns
\bibitem{HemStuXiZh14}
E. Hemsing, G. Stupakov, D. Xiang, A. Zholents,
Beam by design: Laser manipulation of electrons in modern accelerators,
Rev. Mod. Phys. \textbf{86}, 897 (2014).

%higher harmonics, usual bunch, w/o modulator
\bibitem{KatohSRexp}
M. Katoh \textit{et al}.,
Helical phase structure of radiation from an electron in circular motion,
Sci. Rep. \textbf{7}, 6130 (2017).

%2nd harmonic. beam modulation (non helical). helical undulator radiator. sel rules
\bibitem{Rubic17}
P.~R. Ribi\v{c} \textit{et al}.,
Extreme-ultraviolet vortices from a free-electron laser,
Phys. Rev. X \textbf{7}, 031036 (2017).


%tw phtns by bunches. theor%Kaminer16,BKb
%VCh. tw photon radiation by beams of tw e-.
\bibitem{Kaminer16}
I. Kaminer \textit{et al}.,
Quantum \v{C}erenkov radiation: Spectral cutoffs and the role of spin and orbital angular momentum,
Phys. Rev. X \textbf{6}, 011006 (2016).

%tw photons by bunches
\bibitem{BKb}
O.~V. Bogdanov, P.~O. Kazinski,
Probability of radiation of twisted photons by axially symmetric bunches of particles,
arXiv:1811.12616.


%generation of soft tw phtns%KnyzSerb,PadgOAM25,Roadmap16,AndBabAML,TorTorTw,AndrewsSLIA
%обзор (в общие ссылки). twph
\bibitem{KnyzSerb}
B.~A. Knyazev, V.~G. Serbo,
Beams of photons with nonzero projections of orbital angular momenta: New results,
Phys.-Usp. \textbf{61}, 449 (2018).

%gen refs%
\bibitem{PadgOAM25}
M.~J. Padgett,
Orbital angular momentum 25 years on,
Optics Express \textbf{25}, 11267 (2017).

\bibitem{Roadmap16}
H. Rubinsztein-Dunlop \textit{et al}.,
Roadmap on structured light,
J. Opt. \textbf{19}, 013001 (2017).

\bibitem{AndBabAML}
D.~L. Andrews, M. Babiker (Eds.),
\textsl{The Angular Momentum of Light}
(Cambridge University Press, New York, 2013).

\bibitem{TorTorTw}
J.~P. Torres, L. Torner (Eds.),
\textsl{Twisted Photons}
(Wiley-VCH, Weinheim, 2011).

\bibitem{AndrewsSLIA}
D.~L. Andrews (Ed.),
\textsl{Structured Light and Its Applications}
(Academic Press, Amsterdam, 2008).



%direct tw phtns%
%HKDXMHR,BHKMSS,KatohSRexp,Hemsing14hel,HemsingTR12,HemStuXiZh14,Rubic17,BKL2,ABKT,EpJaZo,BKL3,BKL4,HemMar12,HemMarRos11,RibGauNin14,HemRos09,IvSerZay,SasMcNu,TaHaKa,KatohPRL,TairKato,JenSerprl,JenSerepj,Ivanov11,AfanMikh,BordKN
%undulator selection rule
%SasMcNu,BHKMSS,TaHaKa,Rubic17,BKL2
\bibitem{SasMcNu}
S. Sasaki, I. McNulty,
Proposal for generating brilliant X-ray beams carrying orbital angular momentum,
Phys. Rev. Lett. \textbf{100}, 124801 (2008).

%beam manipul. coherent radiation. helical bunches. transition radiation. beam diagnostics. tw phtns
\bibitem{HemRos09}
E. Hemsing, J.~B. Rosenzweig,
Coherent transition radiation from a helically microbunched electron beam,
J. Appl. Phys. \textbf{105}, 093101 (2009).


%beam manipul. by OAM laser. coherent radiation. helical bunches. radiating undulator (with any polarization). theor propn.
\bibitem{HemMarRos11}
E. Hemsing, A. Marinelli, J.~B. Rosenzweig,
Generating optical orbital angular momentum in a high-gain free-electron laser at the first harmonic,
Phys. Rev. Lett. \textbf{106}, 164803 (2011).

%twphtns. mode funcs%JenSerprl,JenSerepj,Ivanov11
\bibitem{JenSerprl}
U.~D. Jentschura, V.~G. Serbo,
Generation of high-energy photons with large orbital angular momentum by Compton backscattering,
Phys. Rev. Lett. \textbf{106}, 013001 (2011).

\bibitem{JenSerepj}
U.~D. Jentschura, V.~G. Serbo,
Compton upconversion of twisted photons: Backscattering of particles with non-planar wave functions,
Eur. Phys. J. C \textbf{71}, 1571 (2011).

\bibitem{Ivanov11}
I.~P. Ivanov,
Colliding particles carrying nonzero orbital angular momentum,
Phys. Rev. D \textbf{83}, 093001 (2011).

%tw rdtn. undul%AfanMikh,BordKN
\bibitem{AfanMikh}
A. Afanasev, A. Mikhailichenko,
On generation of photons carrying orbital angular momentum in the helical undulator,
arXiv:1109.1603.

%tw rdtn. synchtrn
\bibitem{BordKN}
V.~A. Bordovitsyn, O.~A. Konstantinova, E.~A. Nemchenko,
Angular momentum of synchrotron radiation,
Russ. Phys. J. \textbf{55}, 44 (2012).


%beam manipul. coherent radiation. helical bunches. undulators. tw photons. theor propn.
\bibitem{HemMar12}
E. Hemsing, A. Marinelli,
Echo-enabled X-ray vortex generation,
Phys. Rev. Lett. \textbf{109}, 224801 (2012).


%beam manipul. by masked laser. coherent radiation. semihelical bunches. planar radiating undulator (with any polarization).theor propn.
\bibitem{RibGauNin14}
P.~R. Ribi\v{c}, D. Gauthier, G. De Ninno,
Generation of coherent extreme-ultraviolet radiation carrying orbital angular momentum,
Phys. Rev. Lett. \textbf{112}, 203602 (2014).


%VCh by beams of tw e-.
\bibitem{IvSerZay}
I.~P. Ivanov, V.~G. Serbo, V.~A. Zaytsev,
Quantum calculation of the Vavilov-Cherenkov radiation by twisted electrons,
Phys. Rev. A \textbf{93}, 053825 (2016).


%undulator selection rule
%SasMcNu,BHKMSS,TaHaKa,Rubic17,BKL2
\bibitem{TaHaKa}
Y. Taira, T. Hayakawa, M. Katoh,
Gamma-ray vortices from nonlinear inverse Thomson scattering of circularly polarized light,
Sci. Rep. \textbf{7}, 5018 (2017).


%tw photons in lasers%TaHaKa,KatohPRL,TairKato,CLHK,ZYCWS18
%лазеры. пр-ло отбора
\bibitem{KatohPRL}
M. Katoh \textit{et al}.,
Angular momentum of twisted radiation from an electron in spiral motion,
Phys. Rev. Lett. \textbf{118}, 094801 (2017).


%new
\bibitem{BKL2}
O.~V. Bogdanov, P.~O. Kazinski, G.~Yu. Lazarenko,
Probability of radiation of twisted photons by classical currents,
Phys. Rev. A \textbf{97}, 033837 (2018).

%tw.phot. channel:ABKT,EpJaZo
\bibitem{ABKT}
S.~V. Abdrashitov, O.~V. Bogdanov, P.~O. Kazinski, T.~A. Tukhfatullin,
Orbital angular momentum of channeling radiation from relativistic electrons in thin Si crystal,
Phys. Lett. A \textbf{382}, 3141 (2018).

\bibitem{EpJaZo}
V. Epp, J. Janz, M. Zotova,
Angular momentum of radiation at axial channeling,
Nucl. Instrum. Methods B \textbf{436}, 78 (2018).

%coherent tw ph. props. edge rad
\bibitem{BKL3}
O.~V. Bogdanov, P.~O. Kazinski, G.~Yu. Lazarenko,
Probability of radiation of twisted photons in the infrared domain,
Annals Phys. \textbf{406}, 114 (2019).

%tw phtns with recoil.
\bibitem{BKL4}
O.~V. Bogdanov, P.~O. Kazinski, G.~Yu. Lazarenko,
Semiclassical probability of radiation of twisted photons in the ultrarelativistic limit,
arXiv:1903.04024.


%tw photons in lasers%TaHaKa,KatohPRL,TairKato
%лазеры. пр-ло отбора
\bibitem{TairKato}
Y. Taira, M. Katoh,
Generation of optical vortices by nonlinear inverse Thomson scattering at arbitrary angle interactions,
Astrophys. J. \textbf{860}, 45 (2018).


%helical beams%RibGauNin14,HemMar12,HemMarRos11,HemRos09,HemStuXiZh14,HKDXMHR,HemsingTR12

%tw phtns. applctns. rel. new%MHSSF13,PesFriSur15,AfSeSol18,SGACSSK
\bibitem{MHSSF13}
O. Matula, A.~G. Hayrapetyan, V.~G. Serbo, A. Surzhykov, S. Fritzsche,
Atomic ionization of hydrogen-like ions by twisted photons: angular distribution of emitted electrons,
J. Phys. B: At. Mol. Opt. Phys. \textbf{46}, 205002 (2013).

%wps
\bibitem{PesFriSur15}
A.~A. Peshkov, S. Fritzsche, A. Surzhykov,
Ionization of H${}^+_2$ molecular ions by twisted Bessel light,
Phys. Rev. A \textbf{92}, 043415 (2015).

\bibitem{AfSeSol18}
A. Afanasev, V.~G. Serbo, M. Solyanik,
Radiative capture of cold neutrons by protons and deuteron photodisintegration with twisted beams,
J. Phys. G: Nucl. Part. Phys. \textbf{45}, 055102 (2018).

\bibitem{SGACSSK}
M. Solyanik-Gorgone, A. Afanasev, C.~E. Carlson, C.~T. Schmiegelow, F. Schmidt-Kaler,
Excitation of E1-forbidden atomic transitions with electric, magnetic, or mixed multipolarity in light fields carrying orbital and spin angular momentum [Invited],
J. Opt. Soc. Am. B \textbf{36}, 565 (2019).


%tw photons. detectors%LPBFAC,BLCBP,SSDFGCY,LavCourPad,RGMMSCFR,PSTP19
\bibitem{PSTP19}
B. Paroli, M. Siano, L. Teruzzi, M.~A.~C. Potenza,
Single-shot measurement of phase and topological properties of orbital angular momentum radiation through asymmetric lateral coherence,
Phys. Rev. Accel. Beams \textbf{22}, 032901 (2019).

\bibitem{LPBFAC}
J. Leach \textit{et al}.,
Measuring the orbital angular momentum of a single photon,
Phys. Rev. Lett. \textbf{88}, 257901 (2002).

\bibitem{BLCBP}
G.~C.~G. Berkhout \textit{et al}.,
Efficient sorting of orbital angular momentum states of light,
Phys. Rev. Lett. \textbf{105}, 153601 (2010).

\bibitem{SSDFGCY}
T. Su \textit{et al}.,
Demonstration of free space coherent optical communication using integrated silicon photonic orbital angular momentum devices,
Opt. Express \textbf{20}, 9396 (2012).

\bibitem{LavCourPad}
M.~P.~J. Lavery, J. Courtial, M.~J. Padgett,
Measurement of light's orbital angular momentum,
in \textsl{The Angular Momentum of Light},
edited by D.~L. Andrews, M. Babiker
(Cambridge University Press, New York, 2013).

\bibitem{RGMMSCFR}
G. Ruffato \textit{et al.},
A compact difractive sorter for high-resolution demultiplexing of orbital angular momentum beams,
Sci. Rep. \textbf{8}, 10248 (2018).


%diagnostics%HemsingTR12,HemRos09,HLarocque16
%beam diagnostics
\bibitem{HLarocque16}
H. Larocque \textit{et al}.,
Nondestructive measurement of orbital angular momentum for an electron beam,
Phys. Rev. Lett. \textbf{117}, 154801 (2016).

%add rules%RibGauNin14,HemMar12,HemMarRos11,HemRos09,HemStuXiZh14,HKDXMHR,HemsingTR12

%beam manipul. for coherent radiation
%RibGauNin14,HemMar12,HemMarRos11,HemRos09,HemStuXiZh14,HKDXMHR,HemsingTR12,EAllaria08,Hemsing14hel,Rubicx17,XiStup09,Hemsing7516,BGarcia16,Stupakov09,TLiu19,AMak19,EASeddon17,Hemsing19
\bibitem{Stupakov09}
G. Stupakov,
Using the beam-echo effect for generation of short-wavelength radiation,
Phys. Rev. Lett. \textbf{102}, 074801 (2009).

\bibitem{XiStup09}
D. Xiang, G. Stupakov,
Echo-enabled harmonic generation free electron laser,
Phys. Rev. ST Accel. Beams \textbf{12}, 030702 (2009).

\bibitem{Hemsing7516}
E. Hemsing \textit{et al}.,
Echo-enabled harmonics up to the 75th order from precisely tailored electron beams,
Nature Phot. \textbf{10}, 512 (2016).

\bibitem{BGarcia16}
B. Garcia \textit{et al}.,
Method to generate a pulse train of few-cycle coherent radiation,
Phys. Rev. Accel. Beams \textbf{19}, 090701 (2016).

\bibitem{EASeddon17}
E.~A. Seddon \textit{et al}.,
Short-wavelength free-electron laser sources and science: a review,
Rep. Prog. Phys. \textbf{80}, 115901 (2017).

\bibitem{TLiu19}
T. Liu \textit{et al}.,
Generation of ultrashort coherent radiation based on a laser plasma accelerator,
J. Synchrotron Rad. \textbf{26}, 311 (2019).

\bibitem{AMak19}
A. Mak \textit{et al}.,
Attosecond single-cycle undulator light: a review,
Rep. Prog. Phys. \textbf{82}, 025901 (2019).

\bibitem{Hemsing19}
E. Hemsing,
Echo-enabled harmonic generation,
in \textsl{Synchrotron Light Sources and Free-Electron Lasers},
edited by E. Jaeschke, S. Khan, J. Schneider, J. Hastings
(Springer, Cham, 2019).

\bibitem{ACurcio19}
A. Curcio \textit{et al}.,
Beam-based sub-THz source at the CERN linac electron accelerator for research facility,
Phys. Rev. Accel. Beams \textbf{22}, 020402 (2019).

\bibitem{PRRibic19}
P.~R. Ribi\v{c} \textit{et al}.,
Coherent soft X-ray pulses from an echo-enabled harmonic generation free-electron laser,
Nature Photonics (2019). https://doi.org/10.1038/s41566-019-0427-1.


%PDG2018. round bunches
\bibitem{RPPPDG2018}
M. Tanabashi \textit{et al.} (Particle Data Group),
Review of particle physics,
Phys. Rev. D \textbf{98}, 030001 (2018).


%selection rules%Rubic17,KatohPRL,TaHaKa,SasMcNu,TairKato,BKL4,BKL2,BKL3,BHKMSS,BKb,Hemsing14hel,KatohSRexp

%ed rad gen. Class% BlNord37,Nords37,AkhBerQED,BolDavRok,WeinbergB.12,AkhShul,Bord.1
\bibitem{BlNord37}
F. Bloch, A. Nordsieck,
Note on the radiation field of the electron,
Phys. Rev. \textbf{52}, 54 (1937).

\bibitem{Nords37}
A. Nordsieck,
The low frequency radiation of a scattered electron,
Phys. Rev. \textbf{52}, 59 (1937).

\bibitem{AkhBerQED}
A.~I. Akhiezer, V.~B. Berestetskii,
\textsl{Quantum Electrodynamics}
(Interscience Publishers, New York, 1965).

\bibitem{BolDavRok}
B.~M. Bolotovskii, V.~A. Davydov, V.~E. Rok,
Radiation of electromagnetic waves on instantaneous change of the state of the radiating system,
UFN \textbf{126}, 311 (1978)
[Sov. Phys. Usp. \textbf{21}, 865 (1978)].


%trans. rad%GinzbThPhAstr,GinzTsyt
\bibitem{GinzbThPhAstr}
V.~L. Ginzburg,
\textsl{Theoretical Physics and Astrophysics}
(Pergamon, London, 1979).

\bibitem{GinzTsyt}
V.~L. Ginzburg, V.~N. Tsytovich,
\textsl{Transition Radiation and Transition Scattering}
(Hilger, Bristol, 1990).

%edge rad. gen refs
\bibitem{WeinbergB.12}
S. Weinberg,
\textsl{The Quantum Theory of Fields} Vol. 1: \textsl{Foundations}
(Cambridge University Press, Cambridge, 1996).

\bibitem{AkhShul}
A.~I. Akhiezer, N.~F. Shulga,
\textsl{High-Energy Electrodynamics in Matter}
(Gordon and Breach, New York, 1996).

%edge rad
%genref
\bibitem{Bord.1}
V.~G. Bagrov, G.~S. Bisnovatyi-Kogan, V.~A. Bordovitsyn, A.~V. Borisov, O.~F. Dorofeev, V.~Ya. Epp, V.~S. Gushchina, V.~C. Zhukovskii,
\textsl{Synchrotron Radiation Theory and its Development}
(World Scientific, Singapore, 1999).


%tw phtns in medium
\bibitem{BKLm}
O.~V. Bogdanov, P.~O. Kazinski, G.~Yu. Lazarenko,
Probability of radiation of twisted photons in the isotropic dispersive medium,
in preparation.


%special func
\bibitem{GrRy}
I.~S. Gradshteyn, I.~M. Ryzhik,
\textsl{Table of Integrals, Series, and Products}
(Acad. Press, Boston, 1994).


%Mellin trans%ParKam01,Wong,KalKaz3
\bibitem{ParKam01}
R.~B. Paris, D. Kaminski,
\textsl{Asymptotics and Mellin-Barnes Integrals}
(Cambridge University Press, New York, 2001).

\bibitem{Wong}
R. Wong,
\textsl{Asymptotic Approximations of Integrals}
(SIAM, Philadelphia, 2001).

\bibitem{KalKaz3}
I.~S. Kalinichenko, P.~O. Kazinski,
High-temperature expansion of the one-loop effective action induced by scalar and Dirac particles,
Eur. Phys. J. C \textbf{77}, 880 (2017).


%spec funcs
\bibitem{PruBryMar2}
A.~P. Prudnikov, Yu.~A. Brychkov, O.~I. Marichev,
\textsl{Integrals and Series: Special Functions}. Vol. 2
(Taylor \& Francis, London, 1998).


%Mellin trans%GSh,ParKam01,Wong,KalKaz3
\bibitem{GSh}
I.~M. Gel'fand, G.~E. Shilov,
\textsl{Generalized Functions}, Vol. I: \textsl{Properties and Operations}
(Academic Press, New York, 1964).


%exp profile wpkts%KarlJHEPwp,BilBer17
\bibitem{BilBer17}
I. Bialynicki-Birula, Z. Bialynicka-Birula,
Relativistic electron wave packets carrying angular momentum,
Phys. Rev. Lett. \textbf{118}, 114801 (2017).

\bibitem{KarlJHEPwp}
D.~V. Karlovets,
Relativistic vortex electrons: paraxial versus non-paraxial regimes,
Phys. Rev. A \textbf{98}, 012137 (2018).


%helical beams. creation%RibGauNin14,HemMar12,HemMarRos11,HemRos09,HemStuXiZh14,HKDXMHR,HemsingTR12,CLiu16,XLZhu18,LBJu16
%helical beams in plasma by OAM lasers
\bibitem{CLiu16}
C. Liu \textit{et al}.,
Generation of gamma-ray beam with orbital angular momentum in the QED regime,
Phys. Plasmas \textbf{23}, 093120 (2016).


%helical beams in plasma by circularly polarized lasers
\bibitem{XLZhu18}
X.-L. Zhu \textit{et al}.,
Generation of GeV positron and $\ga$-photon beams with controllable angular momentum by intense lasers,
New J. Phys. \textbf{20}, 083013 (2018).

%helical beams in plasma by OAM lasers
\bibitem{LBJu16}
L.~B. Ju \textit{et al}.,
Manipulating the topological structure of ultrarelativistic electron beams using Laguerre-Gaussian laser pulse,
New J. Phys. \textbf{20}, 063004 (2018).


%tw phtns at 1st harm.%HKDXMHR


%BK semiclassics
\bibitem{BaKaStrbook}
V.~N. Baier, V.~M. Katkov, V.~M. Strakhovenko,
\textsl{Electromagnetic Processes at High Energies in Oriented Single Crystals}
(World Scientific, Singapore, 1998).


%tw phtns in lasers%TaHaKa,KatohPRL,TairKato

%special func
\bibitem{LandBess}
L.~J. Landau,
Ratios of Bessel functions and roots of $\al J_\nu(x)+xJ'_\nu(x)=0$,
J. Math. Anal. Appl. \textbf{240}, 174 (1999).


%special func
\bibitem{NIST}
F.~W.~J. Olver, D.~W. Lozier, R.~F. Boisvert, C.~W. Clark, eds.
\textsl{NIST Handbook of Mathematical Functions}
(Cambridge University Press, New York, NY, 2010).






\end{thebibliography}
\end{document}